\documentclass[fleqn,usenatbib]{mnras}

\usepackage{newtxtext,newtxmath}

\usepackage[T1]{fontenc}
\usepackage{ae,aecompl}

\usepackage{graphicx}	
\usepackage{amsmath}	
\usepackage{amssymb}	
\usepackage{multicol}
\usepackage[flushleft]{threeparttable}

\usepackage{amsmath} 

\newcommand{\spitzer}{{\it Spitzer}}

\newcommand{\erg}{{\hbox{erg~s$^{-1}$}}}             

\newcommand{\nh}{$N_{\rm{H}}$}

\newcommand{\etal}{{{et~al.}}}
 
\newcommand{\kms}{km\,s$^{-1}$}   

\newcommand{\msun}{{\hbox{M$_{\odot}$}}}
\newcommand{\lsun}{{\hbox{L$_{\odot}$}}}

\newcommand{\msunpyr}{{\hbox{M$_{\odot}\,\rm{yr^{-1}}$}}}

\newcommand{\0}{\phantom{0}}
\newcommand{\SFR}{\hbox{\rm SFR}}
\newcommand{\SFH}{\hbox{\rm SFH}}

\def\x2{$\chi^{2}$}
\usepackage{tikz,xcolor}

\definecolor{lime}{HTML}{A6CE39}
\DeclareRobustCommand{\orcidicon}{%
	\begin{tikzpicture}
	\draw[lime, fill=lime] (0,0) 
	circle [radius=0.16] 
	node[white] {{\fontfamily{qag}\selectfont \tiny ID}};
	\draw[white, fill=white] (-0.0625,0.095) 
	circle [radius=0.007];
	\end{tikzpicture}
	\hspace{-2mm}
}
\foreach \x in {A, ..., Z}{%
	\expandafter\xdef\csname orcid\x\endcsname{\noexpand\href{https://orcid.org/\csname orcidauthor\x\endcsname}{\noexpand\orcidicon}}
}

\title[X-ray luminosity sub-galactic scaling relations]{Sub-galactic scaling relations between X-ray luminosity, star-formation rate, and stellar mass}

\author[K. Kouroumpatzakis \etal]{K. Kouroumpatzakis$^{1,2}$\orcidA{}\thanks{E-mail: kkouroub@physics.uoc.gr},
\author x A. Zezas$^{1,2,3}$\orcidB{},
\author x P. Sell$^{1,2,4}$\orcidC{},
\author x K. Kovlakas$^{1,2}$\orcidD{},
\newauthor
\author x P. Bonfini$^{1,2,5}$\orcidE{},
\author x S. P. Willner$^{3}$\orcidG{},
\author x M. L. N. Ashby$^{3}$\orcidF{},
\author x A. Maragkoudakis$^{6}$\orcidH{},
\newauthor
and \author[K. Kouroumpatzakis \etal] f T. H. Jarrett$^{7}$\orcidI{}\\
$^{1}$Department of Physics, Univ. of Crete, GR-70013 Heraklion, Greece\\
$^{2}$Institute of Astrophysics, FORTH, GR-71110 Heraklion, Greece\\
$^{3}$Center for Astrophysics \textbar\ Harvard \& Smithsonian\\
$^{4}$Department of Astronomy, University of Florida, Gainesville FL 32611, USA\\
$^{5}$Institute for Astronomy, Astrophysics, Space Applications \& Remote Sensing,\\   National Observatory of Athens, P. Penteli, GR-15236 Athens, Greece\\
$^{6}$Department of Physics and Astronomy, University of Western Ontario, London, ON N6A 3K7, Canada\\
$^{7}$Department of Astronomy, University of Cape Town, Rondebosch, South Africa\\
}

\date{Accepted XXX. Received YYY; in original form ZZZ}

\pubyear{2020}

\begin{document}
\label{firstpage}
\pagerange{\pageref{firstpage}--\pageref{lastpage}}
\maketitle

\begin{abstract}
    X-ray luminosity ($L_X$) originating from high-mass X-ray
    binaries (HMXBs) is tightly correlated with the host galaxy's
    star-formation rate (SFR)\null. We explore this connection at
    sub-galactic scales spanning ${\sim}$7\,dex in SFR and 
    ${\sim}$8\,dex in specific SFR (sSFR)\null. There is good
    agreement with established relations down to ${\rm SFR}
    {\simeq} 10^{-3}$\,\msunpyr, below which an excess of X-ray
    luminosity emerges. This excess likely arises from low mass
    X-ray binaries. The intrinsic scatter of the $L_X$--SFR
    relation is constant, not correlated with SFR\null. Different
    star formation indicators scale with $L_X$ in different ways,
    and we attribute the differences to the effect of star
    formation history. The SFR derived from H$\alpha$ shows the
    tightest correlation with X-ray luminosity because H$\alpha$
    emission probes stellar populations with ages similar to HMXB
    formation timescales, but the H$\alpha$-based SFR is reliable
    only for $\rm sSFR{>}10^{-12}$\,\msunpyr/\msun.
\end{abstract}



\section{Introduction}

Star formation throughout cosmic time has transformed the
Universe. Among other things, it has illuminated it and has created
the foundations for more complex forms to exist. When considered on
kpc scales, star formation has shaped the phenomenology of
galaxies. Two of the most fundamental characteristics of
galaxies are the stellar mass ($M_\star$; past star formation) and
the current/recent star formation, measured by the star-formation
rate (SFR). There is a strong correlation between galaxies' stellar
masses and SFRs, i.e., the galactic main sequence
\citep[e.g.,][]{2007ApJ...660L..47N,2007A&A...468...33E,2007ApJ...670..156D}.

Studies on sub-galactic scales can show to what extent local
conditions are responsible for global scaling relations
\citep[e.g.,][]{2017MNRAS.466.1192M,2020arXiv200104479E}. Comparisons
on sub-galactic scales among galaxies of different types,
star-formation histories (SFH), and metallicities show great
differences \citep[e.g.,][]{2014A&A...571A..72B} because
star formation is not homogeneously dispersed in the galactic volume
\citep[e.g.,][]{2020ApJ...888...92L}.

X-rays probe recent and past star-formation activity and are
particularly useful for characterizing star formation in obscured
environments. X-ray binaries (XRBs) in particular provide a means to
quantify the numbers of stellar remnants (neutron stars and black
holes) otherwise hidden from view. XRBs are formed when a donor star
provides mass to a compact object to which it is gravitationally
bound. The mass transfer can be via Roche lobe overflow or stellar
wind, and either way, the accreting mass radiates at X-ray
wavelengths. Donor stars can be high-mass OB stars or low-mass
stars. Based on their donor stars, systems are described as either
high-mass X-ray binaries (HMXBs) or low-mass X-ray binaries
(LMXBs). Collectively, the X-ray emission from all the XRBs hosted in
a galaxy shows strong correlations with galaxy-wide characteristics
such as SFR and stellar mass. Specifically, LMXB X-ray emission
correlates strongly with stellar mass
\citep[e.g.,][]{2004MNRAS.349..146G,2010ApJ...724..559L,
2011ApJ...729...12B,2012A&A...546A..36Z}, and HMXB X-ray emission
correlates with SFR \citep[e.g.,][]{2003MNRAS.339..793G,
2003A&A...399...39R, 2012MNRAS.419.2095M, 2012MNRAS.426.1870M,
2014MNRAS.437.1698M}.

Recently there have been efforts to examine the $L_X$--SFR--$M_\star$
correlations down to sub-galactic scales in the nearby Universe. The
ratio of XRBs' X-ray output to visible luminosity varies significantly
when examined on small physical scales. This is witnessed by
explorations of the X-ray luminosity of individual regions of a few
nearby galaxies \citep[e.g.,][]{2019MNRAS.483..711A} and by
investigations of the X-ray luminosity functions of XRBs associated
with stellar populations of different ages or metallicities
\citep[e.g.,][]{2019ApJS..243....3L}.

A complication in understanding the correlation between XRBs and SFR
is that there are multiple SFR indicators based on different physical
mechanisms. Indicators include 1.4 GHz emission from
sychrotron radiation of relativistic electrons accelerated
in supernovae remnants, absorbed ultraviolet (UV) radiation heating
galactic dust and being re-emitted at 24\,$\mu$m and in the far infrared, UV from
high mass stars' photospheres, emission lines from atomic gases
ionized by OB stars, polycyclic aromatic hydrocarbons (PAHs) emitting
from the surrounding photo-dissociation regions, etc. This results in
differences between the different SFR indicators that multiple
galaxy-wide studies have tried to calibrate \citep[e.g.,
SFRS;][]{2019MNRAS.482..560M}. The different SFR indicators probe
stellar populations of different ages
\citep[e.g.,][]{2012ARA&A..50..531K} with the ones from ionized
atomic gases probing the most recent
\citep[e.g.,][]{2014A&A...571A..72B, 2016A&A...589A.108C}.

X-ray emission is considered an emerging SFR indicator, but the
correlations still suffer from stochastic and calibration
effects. These effects, which are detected in galaxy-wide
correlations, are increased when examined on sub-galactic scales
because star formation is a local event and hence is diluted on the
surface of a galaxy. Theoretical models predict X-ray luminosity
variations from different stellar populations
\citep[e.g.,][]{2001ApJ...554.1035F,2009MNRAS.395L..71M,2010MNRAS.408..234M}. XRB
population synthesis models show that the bulk of the X-ray output
originating from XRBs is short lived (${\le}$20 Myr) because
that the emission from HMXBs is orders of magnitude higher than
that of LMXBs \citep[e.g.,][]{2013ApJ...764...41F}. Therefore, in
order to understand how biases arise in the X-ray luminosity, SFR,
and stellar-mass correlations, it is important to examine the
correlations on sub-galactic scales and with different SFR
indicators.

Sample selection can bias our interpretation and measurement of the
aforementioned correlations. For example,
\citet[][hereafter M14]{2014MNRAS.437.1698M} studied the $L_X$--SFR
scaling relation for a small sample of star-forming
galaxies. \cite{2004MNRAS.349..146G} and \cite{2011ApJ...729...12B}
studied the $L_X$--$M_\star$ relation for samples of early type
galaxies. \cite{2010ApJ...724..559L} introduced an
$L_X$--SFR--$M_\star$ scaling relation that accounts for the
contribution of HMXBs (scaling with SFR) and LMXBs (scaling with
stellar mass) based on samples of local as well as higher-redshift
galaxies. This analysis used a sample of nearby galaxies with a large
range and mix of stellar masses and SFRs.

This paper's goal is to estimate the effect different star-forming
conditions and SFHs (along with the fact that different SFR
indicators probe different time-scales) may induce in the correlation
and to measure the scatter in each case. The paper is organized as
follows: Section \ref{sec:sample_all} describes the sample of
galaxies, the data/observations, and the data reduction. Section
\ref{sec:Sub-galactic analysis} describes how sub-galactic analysis
was performed. The maximum likelihood fits and the results of the
analysis are described in Section \ref{sec:Results}. The results of
the analysis are discussed in Section \ref{sec:Discussion}, and the
summary is in Section \ref{sec:Summary}.

\section{Sample selection and observations}
\label{sec:sample_all} 

\subsection{Sample}
\label{sec:sample} 

Our galaxy sample is based on the Star Formation Reference Survey
\citep[SFRS;][]{2011PASP..123.1011A}. The SFRS is comprised of 369
galaxies that represent all modes of star formation in the local
Universe. They fully cover the 3D space of three fundamental galaxy
properties: the SFR, indicated by the 60 $\mu$m luminosity; the
specific SFR (sSFR), indicated by the $K_S - F_{60}$ colour; and the
dust temperature, indicated by the FIR ($F_{100}/F_{60}$) flux
density ratio. The SFRS benefits from panchromatic coverage of the
electromagnetic spectrum from radio to X-rays, including optical
spectra of the galaxy nuclei \citep{2017MNRAS.466.1192M} and
H$\alpha$ imaging (Kouroumpatzakis\,\etal\,in prep.). The objective
SFRS selection criteria let us put the sample galaxies in context of
the local star-forming galaxy population.

\begin{figure}
\begin{center}
\includegraphics[width=\columnwidth]{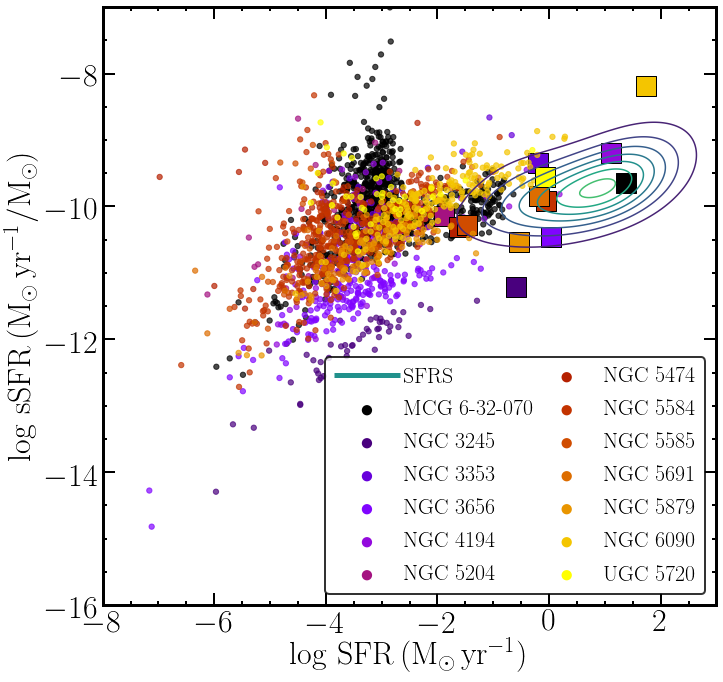}
\caption{The ranges of SFR and sSFR spanned by the sample analysed in
  this work. Small symbols indicate our H$\alpha$-based SFR and
  3.6\,$\mu$m-based stellar mass estimates within $1\times1$\,kpc$^2$
  sub-galactic regions. Squares indicate the integrated emission of
  the sample galaxies. The data are colour-coded by the galaxy they
  refer to. The contours indicate the distribution of the complete
  SFRS.}

    \label{fig:sfr_ssfr_range}
\end{center}
\end{figure}

The sample used for this work consists of 13 star-forming (non-AGN)
SFRS galaxies (Table~\ref{tab:Sample}) for which there are
\textit{Chandra} data of adequate quality to study the X-ray emission
down to 1\,kpc$^2$ scales (Table\,\ref{tab:X-ray_props}) available in
the archive. The sample galaxies span ${\sim}$4\,dex in the total SFR
and ${\sim}$3\,dex in sSFR\null. On sub-galactic scales these ranges
become ${\sim}$7\,dex and ${\sim}$8\,dex in SFR and sSFR respectively
(Fig.~\ref{fig:sfr_ssfr_range}).

\begin{table*}
	\centering
	\caption{Summary of sample galaxies}
	\label{tab:Sample}
	\begin{threeparttable}
	\begin{tabular}{clcccccccc}
     SFRS & Galaxy & Position & D25 & Distance & log\,$L_{60}$ & $K_S-F_{60}$ & $\frac{F_{100}}{F_{60}}$ & Metallicity$^a$ & Axis ratio\\
	 ID & & (J2000) & (\arcsec) & (Mpc) & ($\textit{L}_\odot$) & (AB mag) & & &\\
	\hline
    86& NGC\,3245 & 10:27:18.41 +28:30:26.6 &167&17.8&8.49&1.62&1.60& -&0.52\\
    93& UGC\,5720 & 10:32:31.87 +54:24:03.7 &57&24.9&8.94&5.35&1.15& 8.89$\pm{0.01}$*&0.74\\
    99& NGC\,3353 & 10:45:22.06	+55:57:39.9 &68&18.9&8.67&5.55&1.28& 8.30$\pm{0.01}$*&0.75\\
    124& NGC\,3656 & 11:23:38.64 +53:50:31.7 &97&42.8&9.12&3.27&2.28& -&0.90\\
    182& NGC\,4194 & 12:14:09.65 +54:31:35.9 &92&39.1&10.00&6.04&1.14& 8.88$\pm{0.01}$*&0.65\\
    266& NGC\,5204 & 13:29:36.58 +58:25:13.3 &159&5.1&7.29&4.20&1.76& 8.70$\pm{0.03}$&0.95\\
    300& NGC\,5474 & 14:05:01.42 +53:39:44.4 &54&7.2&7.32&3.70&2.45& 8.80$\pm{0.01}$&0.97\\
    312& NGC\,5585 & 14:19:48.19 +56:43:45.6 &179&8.0&7.33&3.63&2.59& 8.41$\pm{0.01}$*&0.87\\
    314& NGC\,5584 & 14:22:23.76 $-$00:23:15.6 &112&26.7&8.67&3.93&2.39& 8.74$\pm{0.01}$*&0.79\\
    321& MCG\,6-32-070 & 14:35:18.38 +35:07:07.2 &45&126.6&10.07&4.88&2.03& 8.71$\pm{0.02}$&0.95\\
    324& NGC\,5691 & 14:37:53.33 $-$00:23:55.9 &89&30.2&8.90&4.37&1.92& 8.79$\pm{0.01}$&0.61\\
    334& NGC\,5879 & 15:09:46.78 +57:00:00.8 &152&12.4&8.37&3.11&2.76& - &0.93\\
    356& NGC\,6090 & 16:11:40.32 +52:27:23.1 &79&132.4&10.47&6.16&1.54& 8.72$\pm{0.01}$ &0.89\\
	\hline
	 \end{tabular}
    \begin{tablenotes}
            \item ($a$) Metallicities measured using the O3N2 diagnostic \Big(based on $\rm{\log} \frac{[\ion{O}{III}] / H\beta}{[\ion{N}{II}] / H\alpha}$\Big) from \citet{2018MNRAS.475.1485M}. 
            \item (*) Metallicities measured from the galaxy's nucleus.
    \end{tablenotes}
    \end{threeparttable}
\end{table*}

\begin{table*}
    \caption{X-ray data and best-fit model parameters.}
    \centering
    \label{tab:X-ray_props}
    \begin{threeparttable}
    \setlength\tabcolsep{4pt} 
    \begin{tabular}{c|c|c|c|c|c|c|c}
        SFRS & Galaxy & Exp. time & Detector & Spectral Model & $\Gamma$  &  \textit{kT} & \nh \\ 
        ID &       &   (ks)       &          &                &        &  (keV)  & ($10^{22}\rm{cm^{-2}}$) \\
        \hline
        086 & NGC\,3245 & 9.6 & ACIS-S & power-law + APEC  & $2.14_{-0.00}^{+0.08}*$ & $0.52_{-0.22}^{+0.22}$ & $0.02^{+0.08}_{-0.00}*$ \\[3pt]
        093 & UGC\,5720 & 19.2 & ACIS-S & power-law + APEC & $2.25^{+0.23}_{-0.23}$ & $0.83^{+0.01}_{-0.01}$ & $0.01^{+0.01}_{-0.00}*$ \\[3pt]
        099 & NGC\,3353 & 17.8 & ACIS-S & power-law        & $1.42^{+0.18}_{-0.18}$ & &$0.04^{+0.11}_{-0.00}$*\\[3pt]
        124 & NGC\,3656 & 53.8 & ACIS-S & power-law        & $3.60^{+1.00}_{-1.00}$  &  & $0.26^{+0.15}_{-0.15}$ \\[3pt]
        182 & NGC\,4194 & 35.5 & ACIS-S & power-law + APEC & $2.06^{+0.21}_{-0.21}$ & $0.36^{+0.07}_{-0.07}$ & $0.06^{+0.04}_{-0.04}$ \\[3pt]
        266 & NGC\,5204 & 9.8 & ACIS-I & power-law         & $1.68^{+0.10}_{-0.10}$  &  & $0.05^{+0.04}_{-0.04}$\\[3pt]
        300 & NGC\,5474 & 1.7 & ACIS-S & power-law         & $1.02^{+0.05}_{-0.05}$ & &$0.01^{+0.02}_{-0.00}*$\\[3pt]
        312 & NGC\,5585 & 5.3 & ACIS-S & power-law         & $1.46^{+0.43}_{-0.43}$ & &$0.15^{+0.17}_{-0.00}*$\\[3pt]
        314 & NGC\,5584 & 7.0 & ACIS-S & power-law         & $2.41^{+0.40}_{-0.40}$ & &$0.30^{+0.10}_{-0.00}$*\\[3pt]
        321 & MCG\,6-32-070 & 44.6 & ACIS-S & power-law    & $2.42^{+0.91}_{-0.55}$  &  & $0.01^{+0.12}_{-0.00}$*\\[3pt]
        324 & NGC\,5691 & 14.9 & ACIS-S & power-law        & $1.62^{+0.30}_{-0.30}$  &  & $0.06^{+0.08}_{-0.00}$*\\[3pt]
        334 & NGC\,5879 & 89.0 & ACIS-I & power-law        & $1.53^{+0.15}_{-0.15}$  &  & $0.03^{-0.05}_{-0.00}$*\\[3pt]
        356 & NGC\,6090 & 14.8& ACIS-S & power-law  & $3.38^{+0.33}_{-0.33}$  &  & $0.30^{+0.06}_{-0.06}$ \\[3pt]
        \hline
    \end{tabular}
    \begin{tablenotes}
            \item (*) Parameter pegged at the low bound.
    \end{tablenotes}
    \end{threeparttable}
\end{table*}

\subsection{\texorpdfstring{H$\alpha$}{TEXT} data} 
\label{sec:SFR}

The primary SFR indicator used in this work is H$\alpha$ emission,
which traces gas ionized by stellar populations of ages
${\le}$20\,Myr \citep[e.g.,][]{2011ApJ...737...67M}. Because the
formation timescale of HMXBs is typically 10--30\,Myr
\citep[e.g.,][]{2013ApJ...764...41F}, it is in principle well-matched
to SFR probed by H$\alpha$ emission.

We have obtained H$\alpha$ observations with the 1.3 m telescope of
the Skinakas\footnote{http://skinakas.physics.uoc.gr/}
observatory. To account for the redshift range of the SFRS sample
galaxies, we used a custom-built set of filters centered at $\lambda
= 6563, 6595, 6628, 6661, 6694, 6727, 6760$\,\AA\ with
average $\rm FWHM=45$\,\AA. The exposure time for H$\alpha$
observations was 1~hour. We also obtained ${\sim}$10~minute
continuum-band exposures with a filter equivalent to SDSS~$r'$. The
H$\alpha$ observations were taken between 2016 and 2019 under
photometric conditions and typical seeing ${\sim} 1\arcsec$. Details
of the observations and data will be presented by
Kouroumpatzakis \etal\ (in prep).

After the initial reductions (bias subtraction, flat fielding, flux
calibration, etc.) the standard continuum subtraction technique was
performed, based on the relative flux density of the foreground stars
in the continuum and H$\alpha$ images
\citep[e.g.,][]{2008ApJS..178..247K}. This comparison results in a
distribution of $\rm{H{\alpha}/SDSS\,\textit{r}'}$ band flux density
ratios for the various stars included in each frame. We used the mode
of this distribution as the \textit{continuum scaling factor} and its
standard deviation as a measure of the uncertainty of this
procedure. The rescaled continuum image was subtracted from the
H$\alpha$ image to generate the continuum-subtracted H$\alpha$
image. In order to minimize the effect of poorly subtracted stars,
their residuals were masked. These residuals were usually a result of
PSF differences between the narrow band and continuum observations or
colour variations arising from the variety of the foreground stars in
the observed frames.

\begin{figure}
\begin{center}
\includegraphics[width=0.8\linewidth]{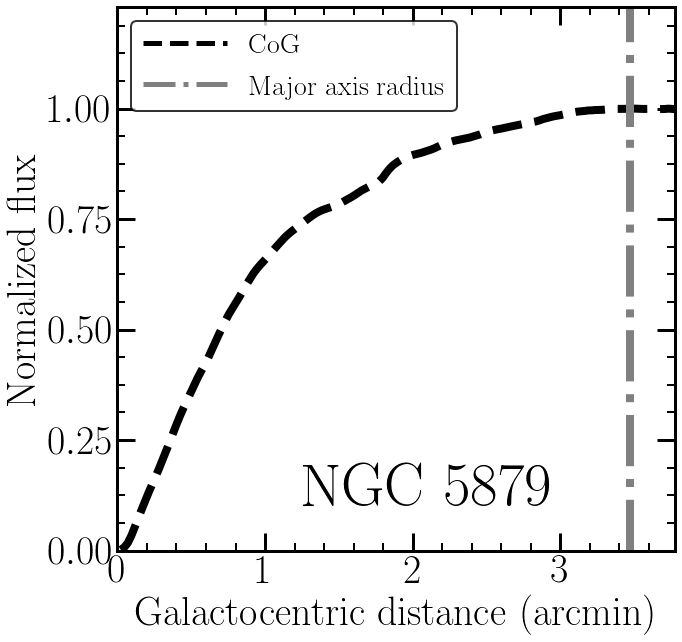}
\caption{Curve of growth (dashed black line) for the H$\alpha$ flux
  of galaxy NGC\,5879. The normalized integrated flux is shown on the
  vertical axis, and the galactocentric distance on the
  horizontal. The semimajor-axis of the aperture, computed following
  the iterative procedure described in the text, is presented by a
  dashed-dotted grey line.}
    \label{fig:CoG}
\end{center}
\end{figure}

A curve of growth (CoG) technique was used to measure the net
H$\alpha$ flux of each galaxy while simultaneously estimating and
subtracting the sky background (Fig.~\ref{fig:CoG}). The background
was estimated by performing a linear fit to the last 5\% of the
CoG\null. This procedure was repeated iteratively while regulating the
background until this part of the CoG was flat. The galaxy aperture
size was defined from the point of the CoG that reaches the
asymptotic line. The aperture shapes used in our analysis were based
on elliptical aperture fits to the \textit{WISE} 4.6\,$\mu$m data of
the SFRS galaxies (following a procedure similar to
\citealt{2019ApJS..245...25J}), keeping the position angle and
ellipticity constant.  The photometric calibrations were based on
observations of spectrophotometric standard stars
\citep{1988ApJ...328..315M}. We included a calibration uncertainty in
our analysis, estimated from the standard deviation of the standard
star's instrumental magnitudes during the observations. The H$\alpha$
luminosity was converted to SFR with the \cite{2011ApJ...737...67M}
conversion:
\begin{eqnarray}
   \frac{ {\rm SFR_{H\alpha}}}{ {\rm (\msun\,  yr^{-1})}} = 10^{-41.27}
    \frac{ {L}_{H\alpha}  }{{\rm ({erg\,  s^{-1}})}} 
	\label{eq:SFR-Ha}
\end{eqnarray}

\subsection{Infrared data} 

In addition to the H$\alpha$ SFR measure, we used \textit{Spitzer}
IRAC non-stellar 8\,$\mu$m and MIPS 24\,$\mu$m observations
\citep{2011PASP..123.1011A}. 8\,$\mu$m probes PAH emission, including
dust-enshrouded star formation. 24\,$\mu$m observations probe warm dust heated by
UV emission from young stars. These two indicators trace
star formation at longer timescales than H$\alpha$ emission
\citep[e.g.,][]{2004ApJ...613..986P,2009ApJ...692..556R,2012ARA&A..50..531K}. The
annuli used for the 8\,$\mu$m and MIPS 24\,$\mu$m analysis were the
same as the H$\alpha$ ones. The background was subtracted as measured
by an annulus outside the galaxy aperture, accounting for any
contribution from foreground stars or background AGN. In the case of
the IRAC 8\,$\mu$m, the stellar continuum was subtracted by rescaling
the 3.6\,$\mu$m images, using the formula from
\cite{2004ApJS..154..253H}:
\begin{eqnarray}
    {f_{8 \micron,{\rm{PAH}}} = f_{8 \micron} - 0.26 \, f_{3.6 \micron}}\quad.
	\label{eq:Helou} 
\end{eqnarray}
Then the non-stellar 8\,$\mu$m luminosity was converted to SFR using
the calibration of \cite{2010ApJ...723..530P}:
\begin{eqnarray}
    \frac{{\rm SFR}_{8\micron,{\rm PAH}}}{(\msunpyr)} = 
    6.3 \times 10^{-10}\frac{{L}_{8 \mu m}}{\lsun}\quad.
	\label{eq:SFR_8}
\end{eqnarray}

The MIPS 24\,$\mu$m luminosity was converted to SFR using the
calibration of \cite{2009ApJ...692..556R}:.
\begin{eqnarray}
    \frac{\rm SFR_{24\micron}}{(\msunpyr)} = 10^{-42.69}
    \frac{{L}_{24\micron}}{ (\erg) }\quad.
	\label{eq:SFR-24}
\end{eqnarray}

The IRAC 3.6\,$\mu$m observations were used to estimate total stellar
masses. The observed flux density was converted to stellar mass using
the \citet{2010RAA....10..329Z} mass-to-light ratio calibration.
\begin{eqnarray}
    \frac{M_{\star}}{\msun} = 10^{0.23+1.14(g-r)}
    \frac{\nu{L}_{\nu 3.6\micron}}{\lsun} \quad.
	\label{eq:M_star}
\end{eqnarray}
where $g$ and $r$ are total galaxy Petrosian AB
magnitudes from SDSS DR12 \citep{2015ApJS..219...12A}. We used each
galaxy's integrated emission $g-r$ colour for all of its
sub-galactic regions. 

\subsection{X-ray Data}
\label{sec:X-rays} 

The \textit{Chandra} data were reduced with \texttt{CIAO v.4.9} and
\textit{CALDB v.4.7.3}. The raw data were reprocessed in order to
apply the latest calibrations and screened for background
flares. Then from the clean event files, we extracted images in the
full (F: 0.5--8\,keV), soft (S: 0.5--2\,keV), and hard (H: 2--8\,keV)
bands and calculated the corresponding monochromatic exposure maps
(at energies of 3.8, 1.5, and 3.8~keV respectively).

For each galaxy we also extracted its integrated spectrum using the
\texttt{CIAO} \texttt{dmextract} command. The extraction aperture was
the same as the H$\alpha$ apertures. Corresponding response and
ancillary response files were also calculated with the \texttt{CIAO
  specextract} tool. Background spectra were extracted from
source-free regions within each field. The X-ray spectra were fitted
with spectral models including power-law, thermal plasma
\citep[APEC;][]{2001ApJ...556L..91S}, and when needed, Gaussian
emission-line components. The spectral analysis was performed using
\texttt{Sherpa v.4.9}. The spectra were binned to have at least 20
counts per bin in order to use the $\chi^{2}$ statistic. The best-fit
model parameters for the integrated spectra of each galaxy are
presented in Table~\ref{tab:X-ray_props}. The details of the spectral
analysis will be presented by Sell~\etal\ (in prep.).

The integrated flux of each galaxy was measured by
integrating the best-fit spectral models. In order to account for
uncertainties in the spectral parameters, the
\texttt{sample\_flux~Sherpa} task was used. This task samples model
parameters from the covariance matrix of the best-fit model, and for
each sample it calculates the corresponding model integrated
flux. This yielded the probability density distribution of the model
flux and the corresponding uncertainties on the spectral parameters.
In addition, for each sample of spectral parameters, the expected
number of counts was calculated by folding the model through the
ancillary response function \citep{2001ApJ...548.1010D} of the
corresponding spectrum. The ratio of the model integrated flux to the
estimated source counts yielded the count-rate to flux conversion
factor, while the distribution of this ratio gave the uncertainty of
the conversion factor as a result of the uncertainty in the model
parameters.

\begin{figure*}
\includegraphics[width=\linewidth]{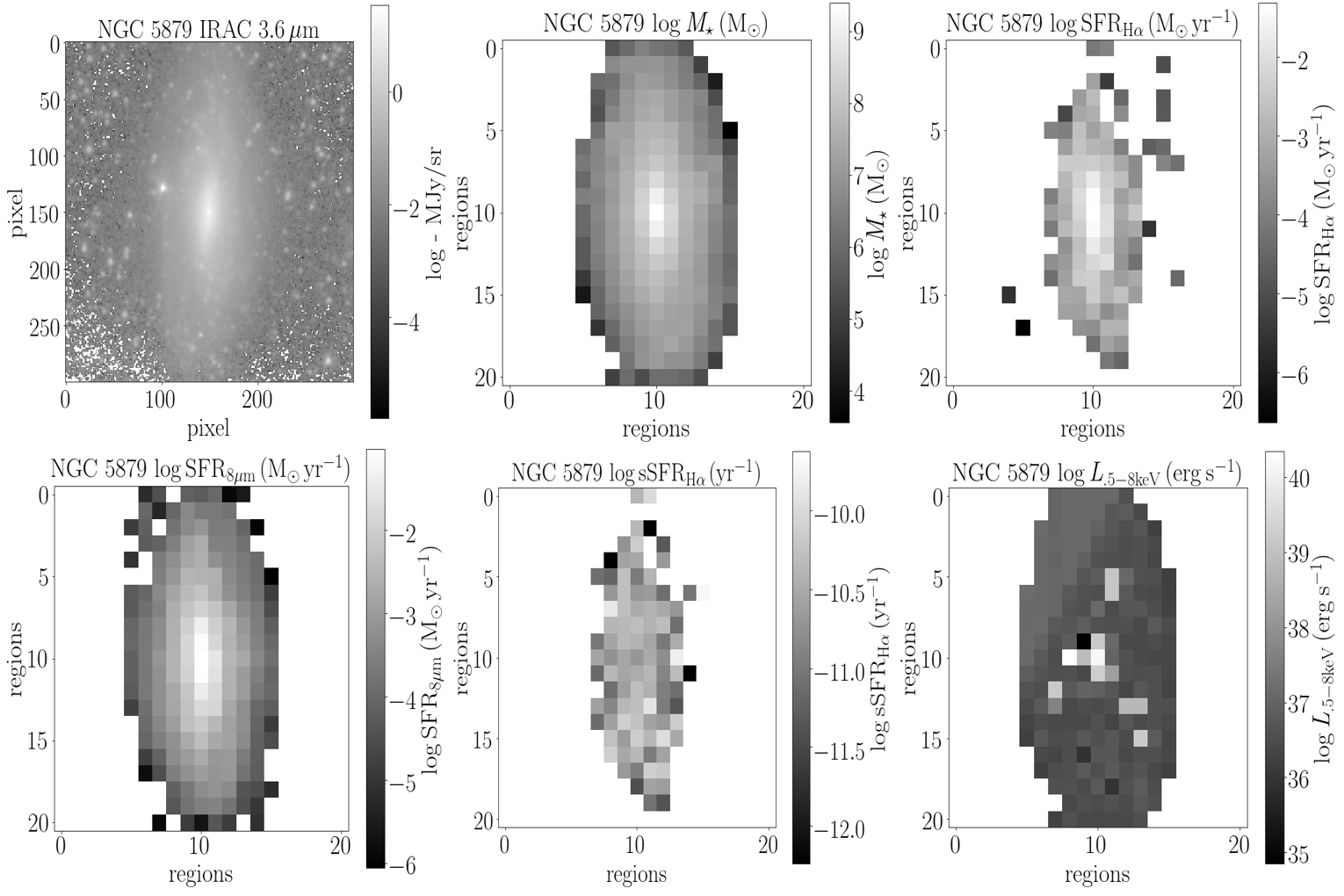}
\caption{Sub-galactic maps at 1\,kpc$^2$ scale for NGC\,5879,
  illustrating the character of the data. Top left: the IRAC
  3.6\,$\mu$m image used to measure the stellar mass.  Top center:
  the stellar mass map derived from the IRAC 3.6\,$\mu$m
  observations. Top right: The SFR map based on the H$\alpha$
  observations. Bottom left: The SFR map derived from the IRAC
  8\,$\mu$m observations. Bottom center: The sSFR map that results
  from combining the H$\alpha$ and the IRAC 3.6\,$\mu$m
  observations. Bottom right: The full (0.5--8\,keV) X-ray luminosity
  map based on the \textit{Chandra} imaging. Bars to the right of
  each image show the mapping from grey scale to physical quantity.}
    \label{fig:Sub_all}
\end{figure*}

\section{Sub-galactic analysis}
\label{sec:Sub-galactic analysis}

In order to explore the correlations between SFR, stellar mass, and
X-ray luminosity on sub-galactic scales, we defined grids of
different physical scales following the same approach as
\citet{2017MNRAS.466.1192M}\null. Physical scales of 1${\times}$1,
2${\times}$2, 3${\times}$3, and 4${\times}$4\,kpc$^2$ were
considered. The minimum physical scale was dictated by the MIPS
24\,$\mu$m PSF (FWHM of centered point spread function =2\farcs6),
which corresponds to a scale of ${\sim}$1\,kpc for the most distant
galaxy (NGC~6090) in our sample (3\farcs14 for 1 kpc regions). One
additional reason for not considering smaller scales is that the SFR
indicators suffer from severe stochasticity at scales ${\lesssim}$1\,kpc \citep[e.g.,][]{2012ARA&A..50..531K}.
Another reason is to ensure that
the natal kicks neutron stars (and possibly black holes) receive will not add significant scatter to the relations
we find. These kicks can result in a considerable velocity
for the surviving binary systems \citep[e.g.,][]{Podsi2004}, 
displacing XRBs from their formation sites.  This could 
increase the scatter in the sub-galactic correlations between SFR and
X-ray luminosity. Typical center-of-mass
velocities measured for HMXBs are in the $15$--$30$\,\kms\ range
\citep[e.g.,][]{2000A&A...364..563V,2005MNRAS.358.1379C,2016MNRAS.459..528A}.
However, for a travel time of
${\sim}20$\,Myr \citep[i.e., the time between formation of the
compact object and the onset of the X-ray emitting phase,
e.g.,][]{2020MNRAS.tmp..540P}, even the upper end of the velocity
range gives a 
distance no more than $\sim$600\,pc from the formation site of an
HMXB\null. In 
the case of LMXBs, their long formation timescales
(${\gtrsim}1$\,Gyr) mean that they trace the old stellar
populations of a galaxy, which are more evenly
distributed. Therefore, the natal kicks will not affect the
statistical association of LMXBs with the older stellar
populations.

We applied the same sub-region grids to all the observables:
IRAC 3.6\,$\mu$m (used to measure
the stellar mass),  H$\alpha$, IRAC 8\,$\mu$m, MIPS 24\,$\mu$m,
(used to measure the SFR), and the \textit{Chandra} data in the soft,
hard, and full bands.
At this stage, regions with
signal-to-noise $\rm (S/N){\le}3$ in the IRAC 3.6\,$\mu$m data were
discarded. This is why the number of sub-galactic regions
does not increase geometrically for smaller physical scales.
The resulting maps of stellar mass, SFR, sSFR,
and X-ray luminosity were used to correlate these parameters in
sub-galactic regions. Figure \ref{fig:Sub_all} shows an example.
Table~\ref{tab:N_regions} lists the number of regions in each of the
galaxies.

\begin{table}
	\centering
	\caption{Number of sub-galactic regions per galaxy at each physical scale.}
	\label{tab:N_regions}
	\begin{tabular}{lrrrr}
\hline\hline
	Galaxy/$\rm{Surface \, (kpc^2)}$ & 1${\times}$1 & 2${\times}$2 & 3${\times}$3 & 4${\times}$4\\
	\hline
    NGC\,3245 & 71 & 25 & 15 & 9\\
    UGC\,5720  &  54 & 20 & 9 & 9\\
    NGC\,3353  &  26 & 10 & 8 & 7\\ 
    NGC\,3656 & 256 & 74 & 40 & 24\\
    NGC\,4194  &  169 & 51 & 29 & 25\\
    NGC\,5204 & 24 & 9 & 8 & 5\\
    NGC\,5474  &  73 & 22 & 11 & 9\\
    NGC\,5585  &  91 & 23 & 17 & 9\\
    NGC\,5584 & 593 & 159 & 80 & 50\\
    MCG\,6-32-070 & 1292 & 337 & 164 & 100\\
    NGC\,5691  &  54 & 19 & 10 & 8\\
    NGC\,5879  &  202 & 62 & 30 & 15\\
    NGC\,6090  &  199 & 57 & 31 & 19\\
	\hline
    Total & 3104 & 868 & 452 & 289\\
	\hline
	\end{tabular}
\end{table}

In order to calculate the X-ray emission in each sub-galactic region,
the observed number of counts was measured using the \texttt{CIAO
dmextract} tool on the \textit{Chandra} images in each of the three
bands. Because most regions had ${\le}$5 counts above the background,
the background could not simply be subtracted as estimated from a
source-free region outside the galaxy. Instead, the
\texttt{BEHR}\footnote{\textit{Bayesian Estimation of Hardness
Ratios};
\url{http://hea-www.harvard.edu/astrostat/BEHR/index.html}}code
\citep{2006ApJ...652..610P} was used, which gives the posterior
probability distribution of the source intensity based on the
formulation of \citet{2001ApJ...548..224V}, accounting for the
Poissonian nature of the source and background counts. \texttt{BEHR}
also takes into account differences in the effective area between the
source and background regions. A non-informative Jeffreys' prior on
the source intensities was adopted. This approach allowed a reliable
estimate of the intensity of the X-ray emission even in regions with
weaker signals than formal detections. It also accounted for
effective area variations across the galaxy's surface based on the
exposure maps of each galaxy.

\begin{figure}
\begin{center}
\includegraphics[width=0.8\linewidth]{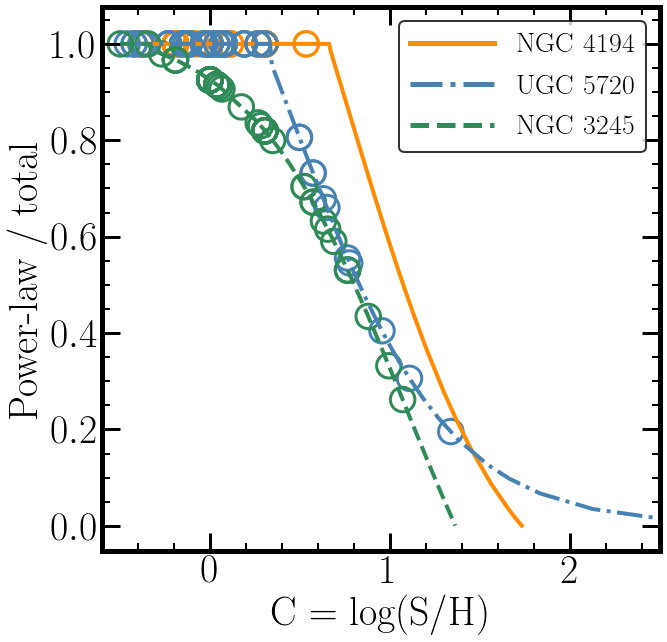}
\caption{The power-law to total ratio as a function of the
  hardness-ratio colour defined as $C \equiv \rm{log}
  \textit{(S/H)}$. These relations were calculated for each galaxy
  independently given the best-fit spectral parameters of their
  integrated spectra and the ACIS calibration. Then, based on each
  region's $C$, we calculated the relative normalization of the
  thermal and power-law component and its uncertainty. Individual
  $\rm{1 {\times} 1 \, kpc^2}$ sub-galactic region measurements are
  represented with circles on top of the calculated conversion
  curves. Colours identify the three galaxies. The cutoff for
  NGC~4194 and NGC~3245 shows that X-ray luminosity for regions
  with $C$ below that value is effectively emitted by a power-law
  spectrum.}
\label{fig:PL_tot_total}
\end{center}
\end{figure}

In order to calculate the X-ray luminosity for each sub-galactic
region, the posterior distribution of the source counts (calculated
as described in Section \ref{sec:X-rays}) was folded with the
distribution of count-rate to flux conversion factors. This
conversion depends on the X-ray spectrum
\citep[e.g.,][]{2006ApJS..166..211Z}. Because each sub-galactic
region has typically ${\le}$20 total counts, no independent spectral
analysis could be performed. Instead the spectrum of each
sub-galactic region was assumed to be the same as the galaxy
integrated spectrum. This is a reasonable assumption for the 10
galaxies fitted with an absorbed power-law spectrum and not requiring
any additional thermal component. The X-ray emission of these
galaxies typically has $\Gamma{\ge} 1.6$ (Section \ref{sec:X-rays},
Table~\ref{tab:X-ray_props}). Therefore their spectra are dominated
by XRBs, which on average have X-ray spectra with photon indices
1${\le}{\Gamma}{\le}$3. The three galaxies that require a thermal
component may have spatial variations in the relative intensity of
the thermal and the power-law components. Assuming that the spectral
parameters of each of the two components are on average the same in
the different sub-galactic regions, the X-ray colour $\rm{\textit{C}
\equiv \log(\textit{S}) - \log(\textit{H})}$ of each region can be
used to infer their relative contribution in the full band. $C$ was
calculated with the \texttt{BEHR}
method. Figure~\ref{fig:PL_tot_total} shows the relation between $C$
and the relative contribution of the power-law to total (power-law +
thermal) components. Based on $C$, the corresponding total flux for
each region and the flux arising only from the power-law component
(which is relevant for the XRBs) were calculated. The mean thermal
contribution for these galaxies is shown in Table~\ref{tab:flux
density_correction}.

\begin{table}
\centering
\caption{Mean (of the distribution of all the sub-galactic regions)
  thermal contribution of the thermal-plasma component in the full
  band $L_{0.5-8\,\rm{keV}}$ luminosity for each galaxy.} 
\label{tab:flux density_correction}
\begin{tabular}{lcccc}
\hline\hline
	Galaxy/$\rm{Surface \, (kpc^2)}$ & 1${\times}$1 & 2${\times}$2 & 3${\times}$3 & 4${\times}$4\\
	\hline
    NGC\,4194 & 0\% & 0\% & 0\% & 0\%\\
    NGC\,3245 & 18.5\% & 19.6\% & 23.1\% & 33.5\%\\
    UGC\,5720 & \08.5\%  & \06.9\% & 16.1\% & \05.8\% \\
\hline
\end{tabular}
\end{table}

The calculation of the X-ray luminosity for each region was performed
by sampling the posterior distribution of the net counts and the
corresponding distribution of count-rate to flux conversions. The
resulting X-ray luminosity distributions are non-Gaussian, usually
positively skewed for low-emission regions.

In order to compare our results with the scaling relations of M14 and
with results from the \textit{Chandra} deep surveys, we also
calculated the luminosities in each sub-galactic region in the soft
and the hard bands. Because the thermal emission included in the soft
band can also be correlated with recent star formation, we opted not
to subtract the thermal component. Therefore the count-rate to flux
conversion factors in the soft band were calculated as described
above, i.e. without correcting for the thermal component. In the case of
the X-ray emission above 2\,keV, which is dominated by the power-law
component, we simply used the best-fit photon index for each galaxy
and its corresponding uncertainty to calculate the distribution of
the count-rate to flux conversion factors.
 
\section{Results}
\label{sec:Results}

\subsection{Maximum likelihood fits}
\label{sec:Fittings}

In order to measure the correlation between X-ray luminosity, SFR,
and stellar mass, we performed maximum likelihood fits using all the
sub-galactic regions of all the galaxies of the sample combined. 
In order to assess the fit parameters and their uncertainties, we
used the posterior probability distribution for the X-ray luminosity,
calculated as described in Section \ref{sec:Sub-galactic analysis},
and the Gaussian uncertainty distribution on SFR and stellar mass of
each region. In all cases, we simultaneously fitted the probability 
distributions of all points included in the fits for all the parameters 
considered in the model. The model is of the form
\begin{eqnarray}
\label{Eq:mdl1} 
\log  L_X =a  \log \SFR + b + \epsilon (\SFR) \quad,
\end{eqnarray} 
where $a$ is the power-law slope
and $b$ the proportionality constant in linear space. We included
an intrinsic scatter term $\epsilon$ to account for any additional
scatter above the measurement random errors. $\epsilon$ is a Gaussian
random variable with mean $\mu = 0$ and standard deviation
$\sigma = \sigma_1  \log\SFR + \sigma_2$. The
intrinsic scatter was allowed to vary linearly (parameterized by $\sigma_1$)
with SFR to account
for stochasticity. This approach was driven by previous studies
\citep[e.g., M14;][]{2003MNRAS.339..793G,2019ApJS..243....3L} which
indicated increased scatter in the $L_X$--SFR scaling relation at
lower SFR\null. The results from these fits are presented in
Table~\ref{tab:Fit_results_2}. In general, slopes are significantly
sub-linear, and $\sigma_1 \simeq 0$ in all fits for all the SFR
indicators and scales used in this work. Thus, even though we are
probing SFRs that extend ${\sim}$5\,dex lower than previous studies,
we do not find significant evidence for increased scatter at lower
SFR\null. Therefore in the rest of our analysis we consider a model with
fixed scatter that does not depend on SFR:
\begin{eqnarray}
\label{Eq:mdl2} 
\log L_X = a \log \SFR + b + \sigma \quad,
\end{eqnarray} 
where $\sigma$ indicates a Gaussian random variable
with $\mu=0$ and standard deviation $\sigma$. The results are
reported in Table~\ref{tab:Fit_results} and described in
Section~\ref{sec:LX_SFR_Correlations}.

In order to disentangle the contribution of HMXBs and LMXBs in the
X-ray luminosity of the sub-galactic regions, we performed a joint
X-ray luminosity, SFR, and stellar mass maximum likelihood fit. The
model was parameterized as
\begin{eqnarray}
\label{Eq:mdl3} 
\log L_X =\log (10^{\alpha + \log{\rm SFR}} + 10^{\beta + \log M_\star})
  + \sigma \quad,
\end{eqnarray} 
where $\alpha$ and $\beta$ are the scaling factors of
the X-ray luminosity resulting from the young and the old stellar
populations (associated with HMXBs and LMXBs respectively), and
$\sigma$ is again a Gaussian random variable accounting for intrinsic
scatter in the data. The fit results are given in 
Table~\ref{tab:Fit_results_2D} and described in 
Section~\ref{sec:LX_SFR_sSFR_Correlations}. The implementation of the maximum
likelihood method is described in more detail in
Appendix~\ref{sec:Apendix_Likelihood}.

\subsection{Correlations between X-ray luminosity and SFR}
\label{sec:LX_SFR_Correlations}

\begin{figure*}
\includegraphics[width=\textwidth]{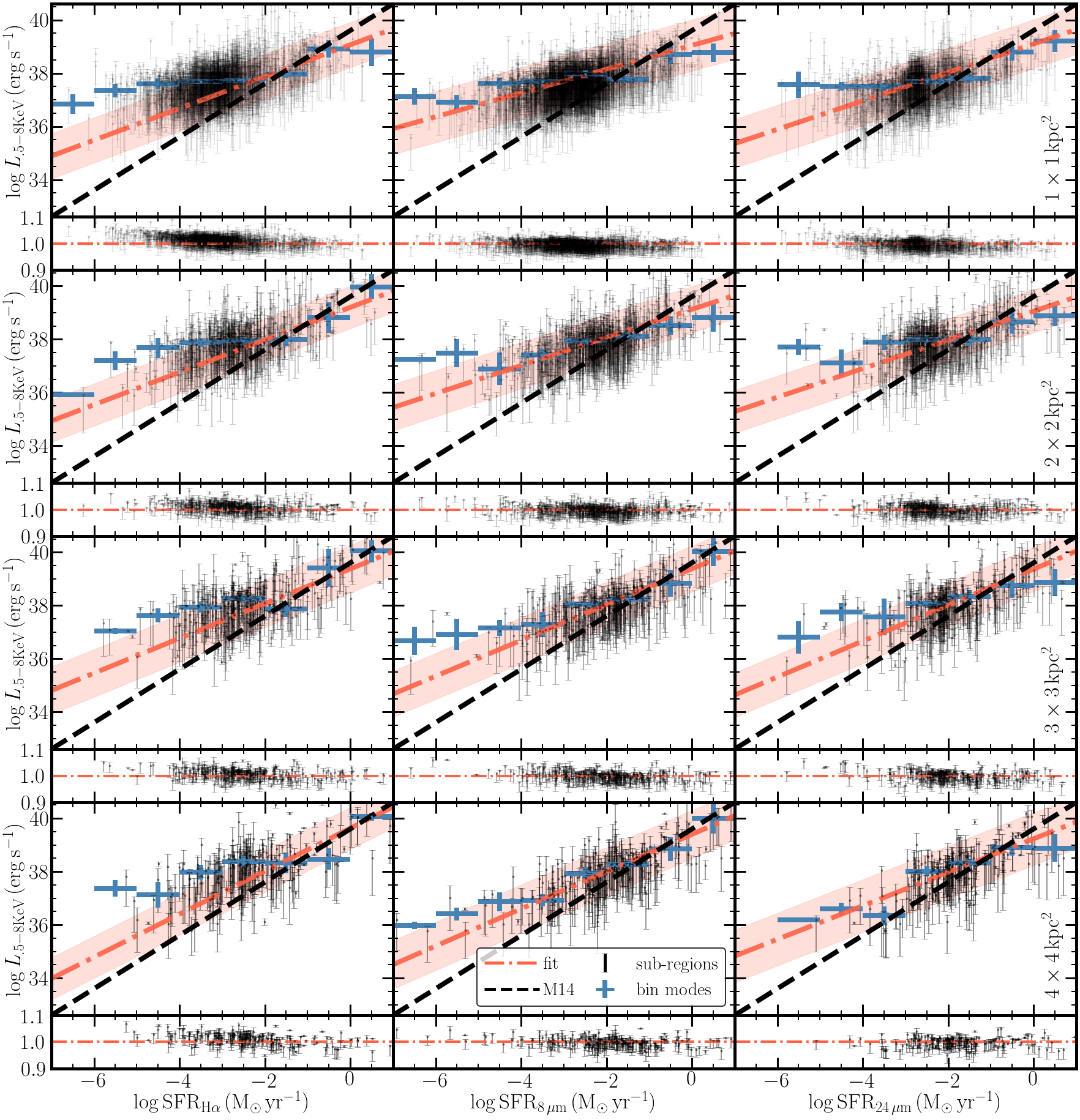}
\caption{$L_{X,0.5-8\rm{keV}}$ as a function of SFR for three
  different SFR indicators (H$\alpha$, 8\,$\mu$m, and 24\,$\mu$m from
  left to right) and for four different sub-galactic scales
  (1${\times}$1, 2${\times}$2, 3${\times}$3, and 4${\times}$4 kpc$^2$
  from top to bottom). All regions within all the sample galaxies are
  included in the fits and are represented by black error bars
  (including uncertainties only in the X-ray luminosity for
  clarity). The red dashed-dotted line represents the maximum
  likelihood best fit for $\log L_X = a \log \SFR + b + \sigma$
  (Eq.~\ref{Eq:mdl2}) for all sub-galactic
  region in the sample. Parameters $a$, $b$, and $\sigma$ are given
  in Table~\ref{tab:Fit_results}. The shaded area represents the
  estimates for the intrinsic scatter $\sigma$. The blue error bars
  represent mode values of the distributions of points included in
  bins of 1 dex of SFR\null. The M14 correlation is drawn with a dashed
  black line. Underneath each panel, the black error bars represent
  for each sub-galactic region
  the ratio of the measured $L_X$ to the value expected based on the
  best-fit model (red dashed-dotted line).}
        \label{fig:Fit_Lx_SFR_all}
\end{figure*}

Figure \ref{fig:Fit_Lx_SFR_all} presents the correlations between
X-ray luminosity and SFR using the H$\alpha$, 8\,$\mu$m, and
24\,$\mu$m SFR indicators. Our analysis used the observed
H$\alpha$ and X-ray luminosities, i.e., not corrected for
absorption. This is because we are interested in deriving empirical
relations between observable quantities. The sample galaxies
show small inclinations (minimum minor-to-major
axis $\rm ratio=0.52$, $\rm
median=0.87\pm0.14$---Table~\ref{tab:Sample}), suggesting 
low intrinsic
absorption. The median extinction for these thirteen galaxies
\citep{2018MNRAS.475.1485M} is $0.36$\,mag based on their integrated
or nuclear spectra. Translating the typical $A_V$-to-hydrogen column
density conversion 
$N_{\rm H}/A_V=1.9 \times 10^{21}$\,atoms\,cm$^{-2}$\,mag$^{-1}$
(\citealt{1975ApJ...198...95G} with cross sections from
\citealt{1983ApJ...270..119M})  to H$\alpha$ with a
\citet{1989ApJ...345..245C} extinction
curve (with $R_V=3.1$) gives
$N_{\rm H}/A_{\rm H\alpha}=1.55 \times
10^{21}$\,atoms\,cm$^{-2}$\,mag$^{-1}$.
This makes the absorption in H$\alpha$ and at 1\,keV 
similar within ${\sim}30$\%.

The best-fit $L_x$--SFR results are presented in
Table~\ref{tab:Fit_results}. Overall correlations between these two
quantities are flatter than the reference correlation of M14. There
are also differences in the slopes depending on the star-formation
indicator considered: H$\alpha$-based SFR shows systematically
steeper slopes, while correlations on the 8\,$\mu$m-based and the
24\,$\mu$m-based emission show shallower slopes. There are also
systematic trends depending on the spatial scales considered. While
the correlations are shallower than linear in all cases, larger
spatial scales tend towards linearity. The shallower slopes are
mainly driven by regions in the extremely low SFR regime, which show
an X-ray luminosity excess in comparison to the linear relation of
M14 and the best maximum likelihood fits from this work.

\begin{figure*}
\includegraphics[width=\textwidth]{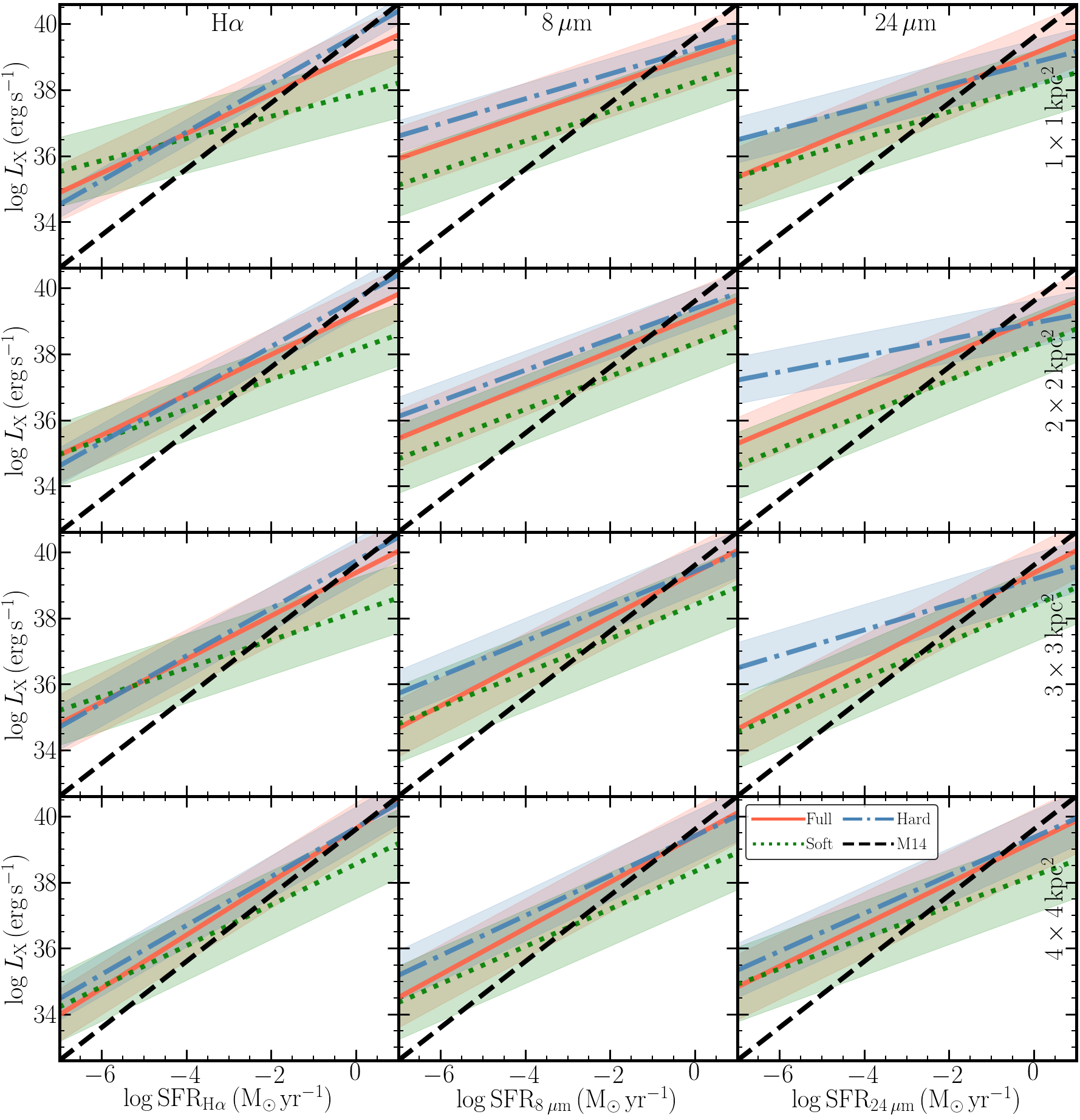}
\caption{Best maximum-likelihood fits to the $L_{X}$--SFR relations
  (see Table~\ref{tab:Fit_results}). The different lines correspond
  to fits for the soft (S; 0.5--2 keV; green dotted), hard (H; 2--8
  keV; blue dashed-dotted), and full (F; 0.5--8 keV; red) bands. Fits
  for the different SFR indicators (H$\alpha$, 8\,$\mu$m, and
  24\,$\mu$m) are shown in the columns from left to right at four
  sub-galactic scales (1$\times$1, 2$\times$2, 3$\times$3, and
  4$\times$4\,kpc$^2$) from top to bottom. The shaded areas of
  similar colours represent the intrinsic scatter $\sigma$ for each
  band. For comparison the M14 correlation is drawn with a black
  dashed line in all panels.}
    \label{fig:Fit_Lx_SFR_TSH}
\end{figure*}

The fits discussed above are based on the full band X-ray data, which
provide the maximum S/N ratio for each sub-region. However, full band
fluxes can be subject to differential absorption and residual thermal
emission. In order to address the importance of these we also
calculated the $L_{X}$--SFR scaling relations in the soft and hard
bands, the latter being a cleaner probe of the X-ray emission
produced by XRBs. The results are presented in 
Table~\ref{tab:Fit_results} and illustrated in 
Figure~\ref{fig:Fit_Lx_SFR_TSH}. The soft band shows weaker correlation with
SFR in all cases. The hard-band fits have similar slopes to the full
band, a fact that reinforces the usefulness of the full band
$L_{X}$--SFR correlation on sub-galactic scales despite the potential
complication of differential absorption. The hard band shows
significantly lower scatter than the full and the soft band in all
cases. The hard band--H$\alpha$ correlation shows the tightest
correlation and slopes closest to one. Especially in the case of the
H$\alpha$-based relation, we find remarkably similar results between
the hard and the full bands. In the case of 2${\times}$2 and
3${\times}$3\,kpc$^2$ 24\,$\mu$m fits, the hard band shows a
shallower fit than the full and soft band. This is due to the
rejection of the low-S/N regions in the 24\,$\mu$m MIPS data. These
regions have very low SFR, reducing the range of SFR and causing the
low-SFR locus to be less populated, thereby driving the flatter fits.

\begin{figure*}
\includegraphics[width=0.99\textwidth]{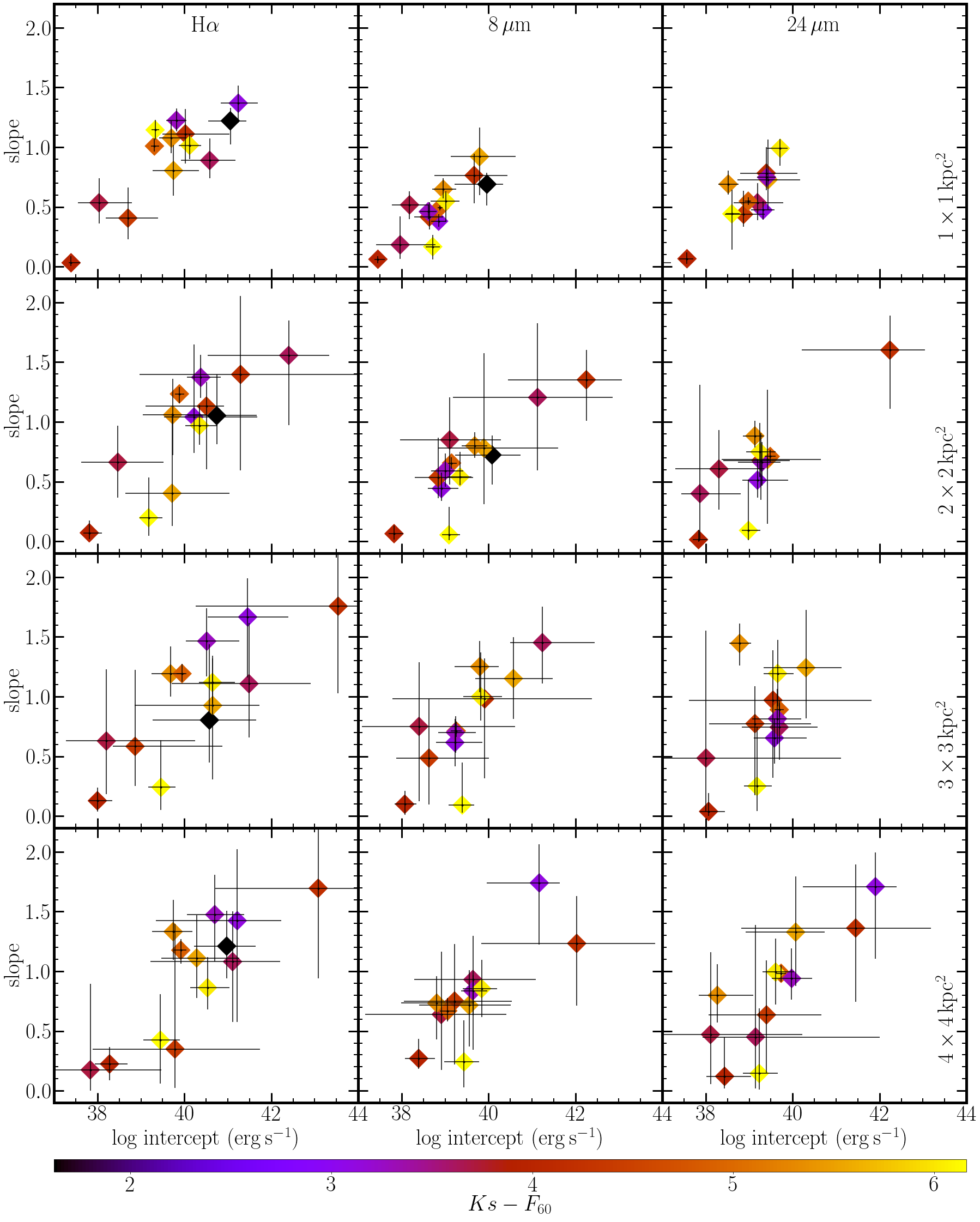}
\caption{Best-fit slopes and intercepts of the sub-regions in each
  individual galaxy. SFR indicators (H$\alpha$, 8\,$\mu$m, and
  24\,$\mu$m) are in columns from left to right, and four
  sub-galactic scales (1${\times}$1, 2${\times}$2, 3${\times}$3, and
  4${\times}$4 kpc$^2$) are from top to bottom. The points are
  colour-coded based on each galaxy's integrated emission $K_S -
  F_{60}$ colour, a proxy for their sSFR\null.}
    \label{fig:Lx_SFR_separ}
\end{figure*}

In order to explore galaxy-to-galaxy variations of the scaling
relations, the model described by Eq.~\ref{Eq:mdl2} was fitted to
each individual galaxy of our sample. The best-fit slopes and
intercepts for the fits for each sub-galactic scale and SFR indicator
are plotted in Figure \ref{fig:Lx_SFR_separ}. We see a broad range of
intercepts and slopes, with some galaxies showing no correlation
(slope${\simeq}$0) and others having slope steeper than 1. As
expected, there is significant correlation between the best-fit slopes
and intercepts. The best-fit parameters for most cases show large
uncertainties (${\simeq}$1 dex) as result of the small number of
regions (Table~\ref{tab:N_regions}) used to derive each
correlation. This is particularly evident as we consider increasing
spatial scales. However, we do see significant differences
between the best-fit slopes and intercepts for the different
galaxies, particularly in the case of the smaller physical scales,
where differences are not masked by large uncertainties. These
variations illustrate the stochasticity in the $L_X$--SFR
correlation, arising from the differences in the SFHs and stellar
populations of the galaxies.

\subsection{Joint correlations between X-ray luminosity, SFR and stellar mass}
\label{sec:LX_SFR_sSFR_Correlations}

\begin{figure*}
\includegraphics[width=\textwidth]{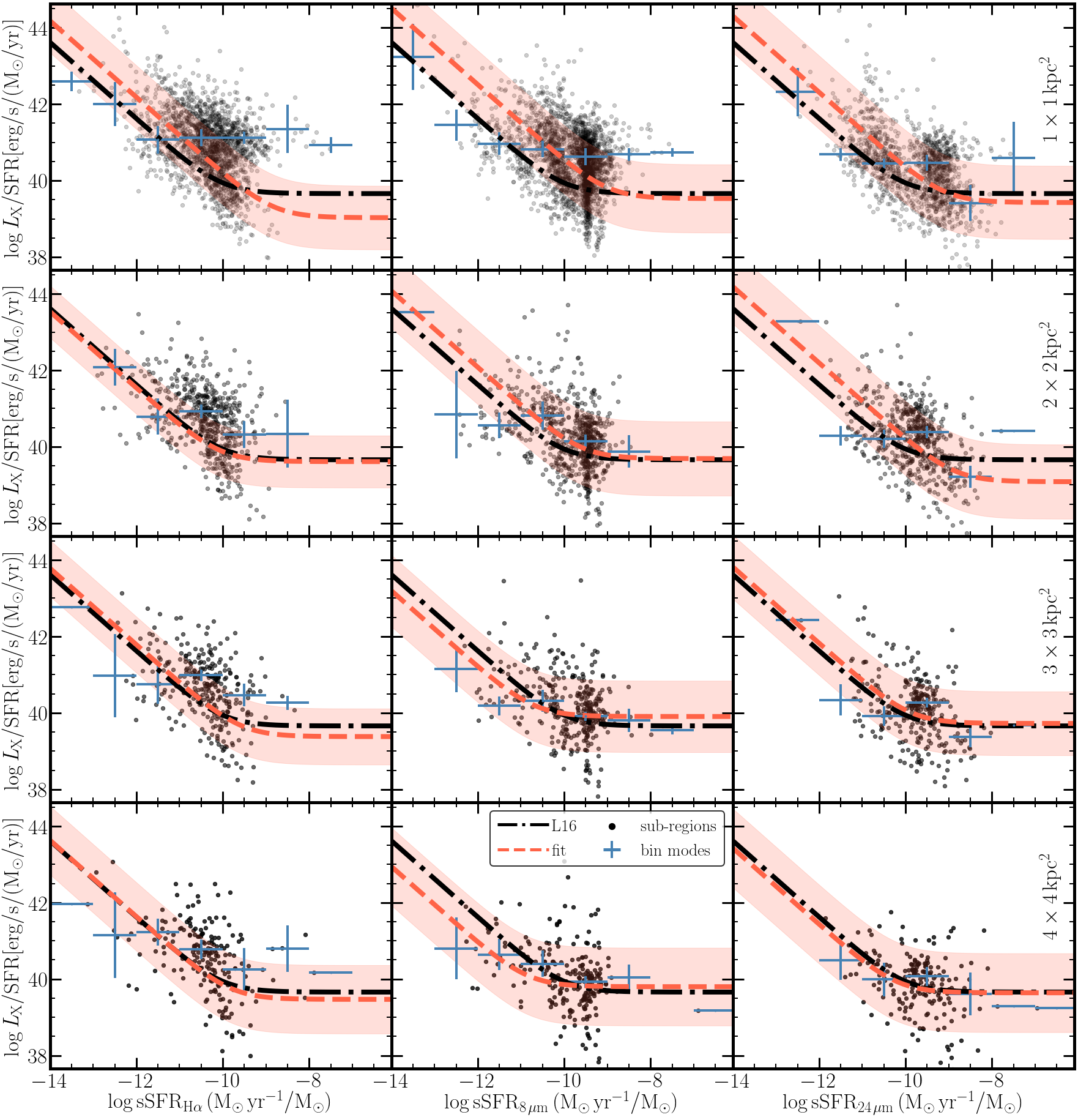}
\caption{$L_{X,0.5-8 \rm{keV}}/\rm{SFR}$ as a function of sSFR with
  the use of three different SFR indicators (H$\alpha$, 8\,$\mu$m,
  and 24\,$\mu$m from left to right) and for four different
  sub-galactic scales (1${\times}$1, 2${\times}$2, 3${\times}$3, and
  4${\times}$4 kpc$^2$ from top to bottom). All regions of all the
  sample galaxies are represented by grey points. The red dashed
  curve represents the best fit for a
$\log L_X =\log (10^{\alpha + \log{\rm SFR}} + 10^{\beta + \log M_\star})
  + \sigma$ 
  model  (Eq.~\ref{Eq:mdl3}). The shaded area represents the
  calculated intrinsic scatter $\sigma$. The blue error bars
  represent the modes and 1$\sigma$ uncertainties of the
  distributions of points in 1 dex bins of sSFR\null. The L16 relation for
  zero redshift is plotted with a black dashed-dotted curve.}
    \label{fig:LxSFR_sSFR}
\end{figure*}

The sSFR is a metric of the relative contribution of the young and
old stellar populations in the mass assembly of the galaxy. Because
HMXBs are associated with young, and LMXBs with old stellar
populations, the sSFR is a proxy for the relative contribution of
these two XRB populations in the overall X-ray emission of a
galaxy. Figure \ref{fig:LxSFR_sSFR} illustrates these correlations
projected on the $L_X$--SFR--$M_\star$ plane. For almost all cases,
we find excellent agreement with the $z {<} 0.5$
(\citealt{2016ApJ...825....7L}; hereafter L16) relation for the
integrated properties of galaxies, even though the results presented
here consider sub-galactic scales and extend these relations to
${\sim}$2 dex lower sSFR\null. The agreement is better for larger scales,
with smaller scales tending to give larger $\alpha$ (Eq. \ref{Eq:mdl3}). As in the case
for the $L_X$--SFR correlations, the scatter is smallest for the
H$\alpha$ SFR indicator.

\begin{table*}
    \centering
    \caption{Maximum likelihood fits of the $L_X$--SFR relation with SFR-dependent scatter.}
        \label{tab:Fit_results_2}
    \begin{threeparttable}
    \begin{tabular}{@{\hspace{0pt}}
                    r@{\hspace{10pt}}
                    c@{\hspace{6pt}}
                    c@{\hspace{6pt}}
                    c@{\hspace{6pt}}
                    c@{\hspace{6pt}}
                    c@{\hspace{6pt}}
                    c@{\hspace{6pt}}
                    c@{\hspace{6pt}}
                    c@{\hspace{6pt}}
                    c@{\hspace{6pt}}
                    c@{\hspace{6pt}}
                    c@{\hspace{6pt}}
                    c@{\hspace{0pt}}}
        Scale & \multicolumn{4}{c}{H$\alpha$}
              & \multicolumn{4}{c}{8\,$\mu$m}
              & \multicolumn{4}{c}{24\,$\mu$m}
            \\
            & $a$ & $b$ & $\sigma_1$ & $\sigma_2$
            & $a$ & $b$ & $\sigma_1$ & $\sigma_2$
            & $a$ & $b$ & $\sigma_1$ & $\sigma_2$
            \\\hline
        1${\times}$1\,kpc$^2$ & $0.50^{+0.02}_{-0.02}$ & $38.87^{+0.03}_{-0.05}$ & $0.14^{+0.01}_{-0.01}$ & $1.13^{+0.04}_{-0.03}$ & $0.44^{+0.01}_{-0.02}$ & $38.99^{+0.05}_{-0.05}$ & $0.18^{+0.02}_{-0.02}$ & $1.43^{+0.05}_{-0.05}$ & $0.46^{+0.03}_{-0.02}$ & $38.87^{+0.09}_{-0.05}$ & $0.14^{+0.02}_{-0.02}$ & $1.27^{+0.04}_{-0.06}$ \\[2pt]
        2${\times}$2\,kpc$^2$ & $0.60^{+0.05}_{-0.05}$ & $39.15^{+0.15}_{-0.09}$ & $0.04^{+0.04}_{-0.03}$ & $0.87^{+0.10}_{-0.06}$ & $0.51^{+0.03}_{-0.04}$ & $39.11^{+0.07}_{-0.11}$ & $0.01^{+0.04}_{-0.02}$ & $0.91^{+0.07}_{-0.05}$ & $0.46^{+0.07}_{-0.02}$ & $38.93^{+0.18}_{-0.05}$ & $0.11^{+0.04}_{-0.05}$ & $1.10^{+0.09}_{-0.14}$ \\[2pt]
        3${\times}$3\,kpc$^2$ & $0.66^{+0.09}_{-0.05}$ & $39.42^{+0.15}_{-0.16}$ & $-0.03^{+0.08}_{-0.04}$ & $0.87^{+0.15}_{-0.12}$ & $0.63^{+0.05}_{-0.05}$ & $39.26^{+0.10}_{-0.14}$ & $0.21^{+0.04}_{-0.04}$ & $1.17^{+0.12}_{-0.06}$ & $0.58^{+0.07}_{-0.04}$ & $39.22^{+0.10}_{-0.13}$ & $0.09^{+0.03}_{-0.04}$ & $1.02^{+0.07}_{-0.09}$ \\[2pt]
        4${\times}$4\,kpc$^2$ & $0.76^{+0.08}_{-0.06}$ & $39.57^{+0.16}_{-0.15}$ & $0.05^{+0.06}_{-0.04}$ & $0.89^{+0.13}_{-0.10}$ & $0.69^{+0.06}_{-0.06}$ & $39.35^{+0.14}_{-0.14}$ & $0.15^{+0.06}_{-0.08}$ & $1.13^{+0.10}_{-0.13}$ & $0.75^{+0.06}_{-0.06}$ & $39.45^{+0.11}_{-0.13}$ & $-0.09^{+0.04}_{-0.04}$ & $0.82^{+0.09}_{-0.08}$ \\[2pt]
    \hline
    \end{tabular}
\begin{tablenotes}
\item NOTE: Model $\rm{\log}\textit{L}_X = \textit{a} \, \rm{\log}SFR
  + \textit{b} + \epsilon (\rm{SFR})$, where $\epsilon$ is a Gaussian
  random variable with mean $\mu = 0$ and standard deviation
  $\rm{\sigma = \sigma_1 \, \rm{\log} SFR + \sigma_2}$ for the full
  (0.5-8 keV) X-ray band.
    \end{tablenotes}
    \end{threeparttable}
\end{table*}

\begin{table*}
\caption{Maximum-likelihood fits of the $L_X$--SFR relation with
    constant scatter.} 
\label{tab:Fit_results}
    \begin{threeparttable}
    \centering
    \begin{tabular}{rccccccccc}
        Scale & \multicolumn{3}{c}{H$\alpha$}
              & \multicolumn{3}{c}{8\,$\mu$m}
              & \multicolumn{3}{c}{24\,$\mu$m}
            \\
            & $a$ & $b$ & $\sigma$
            & $a$ & $b$ & $\sigma$
            & $a$ & $b$ & $\sigma$
\\\hline
\hline    & \multicolumn{9}{c}{Full $L_X$} \\\hline
        1${\times}$1\,kpc$^2$ & $0.60^{+0.01}_{-0.01}$ & $39.07^{+0.03}_{-0.03}$ & $0.85^{+0.01}_{-0.01}$ & $0.45^{+0.02}_{-0.01}$ & $39.04^{+0.05}_{-0.04}$ & $0.96^{+0.01}_{-0.01}$ & $0.54^{+0.01}_{-0.02}$ & $39.10^{+0.04}_{-0.05}$ & $0.92^{+0.02}_{-0.01}$ \\[2pt]
        2${\times}$2\,kpc$^2$ & $0.61^{+0.06}_{-0.01}$ & $39.20^{+0.14}_{-0.04}$ & $0.79^{+0.02}_{-0.02}$ & $0.53^{+0.02}_{-0.04}$ & $39.13^{+0.05}_{-0.08}$ & $0.87^{+0.02}_{-0.02}$ & $0.54^{+0.02}_{-0.04}$ & $39.05^{+0.04}_{-0.09}$ & $0.80^{+0.03}_{-0.02}$ \\[2pt]
        3${\times}$3\,kpc$^2$ & $0.65^{+0.06}_{-0.06}$ & $39.37^{+0.14}_{-0.14}$ & $0.89^{+0.05}_{-0.04}$ & $0.67^{+0.05}_{-0.04}$ & $39.38^{+0.12}_{-0.09}$ & $0.86^{+0.04}_{-0.03}$ & $0.68^{+0.03}_{-0.03}$ & $39.37^{+0.09}_{-0.07}$ & $0.84^{+0.05}_{-0.02}$ \\[2pt]
        4${\times}$4\,kpc$^2$ & $0.81^{+0.07}_{-0.05}$ & $39.63^{+0.15}_{-0.12}$ & $0.79^{+0.04}_{-0.05}$ & $0.70^{+0.06}_{-0.06}$ & $39.40^{+0.11}_{-0.15}$ & $0.91^{+0.05}_{-0.04}$ & $0.63^{+0.05}_{-0.04}$ & $39.23^{+0.11}_{-0.09}$ & $0.99^{+0.05}_{-0.04}$ \\[2pt]
    \hline    & \multicolumn{9}{c}{Soft $L_X$} \\\hline
        1${\times}$1\,kpc$^2$ & $0.34^{+0.03}_{-0.02}$ & $37.87^{+0.07}_{-0.04}$ & $1.05^{+0.01}_{-0.01}$ & $0.45^{+0.01}_{-0.01}$ & $38.24^{+0.03}_{-0.03}$ & $0.94^{+0.01}_{-0.01}$ & $0.40^{+0.02}_{-0.01}$ & $38.13^{+0.05}_{-0.05}$ & $1.06^{+0.02}_{-0.01}$ \\[2pt]
        2${\times}$2\,kpc$^2$ & $0.45^{+0.02}_{-0.02}$ & $38.13^{+0.07}_{-0.05}$ & $0.95^{+0.02}_{-0.02}$ & $0.50^{+0.04}_{-0.02}$ & $38.32^{+0.07}_{-0.06}$ & $1.04^{+0.03}_{-0.02}$ & $0.52^{+0.02}_{-0.01}$ & $38.24^{+0.05}_{-0.05}$ & $1.01^{+0.02}_{-0.02}$ \\[2pt]
        3${\times}$3\,kpc$^2$ & $0.43^{+0.03}_{-0.02}$ & $38.18^{+0.12}_{-0.05}$ & $1.05^{+0.03}_{-0.03}$ & $0.52^{+0.06}_{-0.04}$ & $38.41^{+0.08}_{-0.10}$ & $1.16^{+0.05}_{-0.07}$ & $0.55^{+0.04}_{-0.03}$ & $38.38^{+0.09}_{-0.07}$ & $1.10^{+0.04}_{-0.04}$ \\[2pt]
        4${\times}$4\,kpc$^2$ & $0.62^{+0.05}_{-0.07}$ & $38.55^{+0.15}_{-0.10}$ & $1.04^{+0.05}_{-0.05}$ & $0.57^{+0.03}_{-0.11}$ & $38.33^{+0.10}_{-0.11}$ & $1.12^{+0.07}_{-0.04}$ & $0.47^{+0.02}_{-0.07}$ & $38.19^{+0.09}_{-0.11}$ & $1.13^{+0.06}_{-0.04}$ \\[2pt]
    \hline    & \multicolumn{9}{c}{Hard $L_X$} \\\hline
        1${\times}$1\,kpc$^2$ & $0.73^{+0.02}_{-0.02}$ & $39.65^{+0.04}_{-0.04}$ & $0.39^{+0.02}_{-0.01}$ & $0.38^{+0.01}_{-0.02}$ & $39.24^{+0.05}_{-0.05}$ & $0.48^{+0.03}_{-0.02}$ & $0.33^{+0.02}_{-0.03}$ & $38.83^{+0.06}_{-0.08}$ & $0.69^{+0.03}_{-0.02}$ \\[2pt]
        2${\times}$2\,kpc$^2$ & $0.72^{+0.03}_{-0.04}$ & $39.67^{+0.07}_{-0.06}$ & $0.56^{+0.04}_{-0.02}$ & $0.47^{+0.03}_{-0.03}$ & $39.38^{+0.07}_{-0.07}$ & $0.59^{+0.04}_{-0.03}$ & $0.25^{+0.05}_{-0.03}$ & $38.94^{+0.09}_{-0.09}$ & $0.71^{+0.05}_{-0.05}$ \\[2pt]
        3${\times}$3\,kpc$^2$ & $0.72^{+0.06}_{-0.04}$ & $39.73^{+0.08}_{-0.11}$ & $0.68^{+0.05}_{-0.06}$ & $0.53^{+0.05}_{-0.04}$ & $39.43^{+0.09}_{-0.11}$ & $0.70^{+0.06}_{-0.06}$ & $0.38^{+0.07}_{-0.04}$ & $39.18^{+0.14}_{-0.10}$ & $0.78^{+0.06}_{-0.06}$ \\[2pt]
        4${\times}$4\,kpc$^2$ & $0.74^{+0.06}_{-0.04}$ & $39.65^{+0.10}_{-0.08}$ & $0.64^{+0.05}_{-0.06}$ & $0.60^{+0.06}_{-0.05}$ & $39.40^{+0.12}_{-0.09}$ & $0.76^{+0.06}_{-0.06}$ & $0.58^{+0.07}_{-0.06}$ & $39.36^{+0.11}_{-0.15}$ & $0.82^{+0.07}_{-0.07}$ \\[2pt]
    \hline
    \end{tabular}
    \begin{tablenotes}
\item NOTE: Model $\rm{\log} \textit{L}_X = \textit{a} \,
  \rm{\log}SFR + \textit{b} + \sigma$, where $\sigma$ indicates a
  Gaussian random variable with mean $\mu=0$ and standard deviation
  $\sigma$. 
    \end{tablenotes}
    \end{threeparttable}
\end{table*}

\begin{table*}
    \caption{The maximum likelihood fit results for $L_X$--SFR--$M_\star$ relation.}
    \label{tab:Fit_results_2D}
    \centering
    \begin{threeparttable}
    \begin{tabular}{rccccccccc}
        Scale & \multicolumn{3}{c}{H$\alpha$}
              & \multicolumn{3}{c}{8\,$\mu$m}
              & \multicolumn{3}{c}{24\,$\mu$m}
            \\
            & $\alpha$ & $\beta$ & $\sigma$
            & $\alpha$ & $\beta$ & $\sigma$
            & $\alpha$ & $\beta$ & $\sigma$
            \\\hline
    \hline    & \multicolumn{9}{c}{Full $L_X$} \\\hline
        1${\times}$1\,kpc$^2$ & $39.03^{+0.06}_{-0.06}$ & $30.18^{+0.04}_{-0.03}$ & $0.83^{+0.02}_{-0.02}$ & $39.53^{+0.07}_{-0.07}$ & $30.51^{+0.02}_{-0.02}$ & $0.89^{+0.01}_{-0.01}$ & $39.43^{+0.06}_{-0.10}$ & $30.31^{+0.04}_{-0.04}$ & $0.96^{+0.02}_{-0.01}$ \\[2pt]
        2${\times}$2\,kpc$^2$ & $39.61^{+0.05}_{-0.06}$ & $29.54^{+0.06}_{-0.07}$ & $0.68^{+0.03}_{-0.03}$ & $39.69^{+0.08}_{-0.13}$ & $30.08^{+0.10}_{-0.05}$ & $0.97^{+0.02}_{-0.03}$ & $39.09^{+0.43}_{-0.11}$ & $30.19^{+0.06}_{-0.16}$ & $0.97^{+0.03}_{-0.03}$ \\[2pt]
        3${\times}$3\,kpc$^2$ & $39.38^{+0.09}_{-0.22}$ & $29.78^{+0.08}_{-0.07}$ & $0.73^{+0.06}_{-0.05}$ & $39.91^{+0.06}_{-0.07}$ & $29.20^{+0.27}_{-0.26}$ & $0.93^{+0.04}_{-0.02}$ & $39.72^{+0.05}_{-0.15}$ & $29.82^{+0.10}_{-0.25}$ & $0.83^{+0.11}_{-0.01}$ \\[2pt]
        4${\times}$4\,kpc$^2$ & $39.47^{+0.21}_{-0.20}$ & $29.63^{+0.13}_{-0.13}$ & $0.89^{+0.07}_{-0.08}$ & $39.80^{+0.16}_{-0.02}$ & $28.94^{+0.35}_{-0.41}$ & $1.02^{+0.04}_{-0.06}$ & $39.64^{+0.13}_{-0.02}$ & $29.43^{+0.17}_{-0.08}$ & $1.03^{+0.02}_{-0.08}$ \\[2pt]
    \hline    & \multicolumn{9}{c}{Soft $L_X$} \\\hline
        1${\times}$1\,kpc$^2$ & $38.04^{+0.04}_{-0.04}$ & $29.60^{+0.02}_{-0.02}$ & $0.91^{+0.01}_{-0.01}$ & $38.15^{+0.13}_{-0.19}$ & $29.85^{+0.02}_{-0.01}$ & $0.94^{+0.01}_{-0.01}$ & $37.64^{+0.28}_{-0.24}$ & $29.80^{+0.02}_{-0.02}$ & $0.96^{+0.02}_{-0.01}$ \\[2pt]
        2${\times}$2\,kpc$^2$ & $37.58^{+0.35}_{-0.12}$ & $29.37^{+0.04}_{-0.04}$ & $0.95^{+0.03}_{-0.02}$ & $< 37.25$ & $29.52^{+0.04}_{-0.03}$ & $0.98^{+0.03}_{-0.02}$ & $< 38.79$ & $29.45^{+0.05}_{-0.06}$ & $0.99^{+0.04}_{-0.03}$ \\[2pt]
        3${\times}$3\,kpc$^2$ & $< 38.37$ & $29.34^{+0.06}_{-0.06}$ & $1.01^{+0.03}_{-0.05}$ & $< 37.53$ & $29.37^{+0.05}_{-0.04}$ & $0.97^{+0.04}_{-0.04}$ & $< 37.52$ & $29.33^{+0.05}_{-0.04}$ & $0.99^{+0.03}_{-0.04}$ \\[2pt]
        4${\times}$4\,kpc$^2$ & $38.19^{+0.22}_{-0.27}$ & $29.14^{+0.14}_{-0.04}$ & $1.12^{+0.05}_{-0.06}$ & $< 37.87$ & $29.30^{+0.08}_{-0.06}$ & $0.98^{+0.11}_{-0.04}$ & $< 38.55$ & $29.25^{+0.08}_{-0.09}$ & $1.09^{+0.05}_{-0.09}$ \\[2pt]
    \hline    & \multicolumn{9}{c}{Hard $L_X$} \\\hline
        1${\times}$1\,kpc$^2$ & $39.99^{+0.01}_{-0.01}$ & $29.24^{+0.10}_{-0.10}$ & $0.65^{+0.01}_{-0.01}$ & $40.43^{+0.05}_{-0.04}$ & $30.28^{+0.11}_{-0.07}$ & $1.17^{+0.02}_{-0.02}$ & $40.11^{+0.05}_{-0.04}$ & $30.22^{+0.08}_{-0.06}$ & $1.05^{+0.02}_{-0.02}$ \\[2pt]
        2${\times}$2\,kpc$^2$ & $40.00^{+0.05}_{-0.08}$ & $29.10^{+0.22}_{-0.15}$ & $0.76^{+0.06}_{-0.03}$ & $40.05^{+0.07}_{-0.07}$ & $29.80^{+0.21}_{-0.17}$ & $1.13^{+0.03}_{-0.03}$ & $< 38.58$ & $30.62^{+0.06}_{-0.06}$ & $1.07^{+0.04}_{-0.04}$ \\[2pt]
        3${\times}$3\,kpc$^2$ & $39.98^{+0.06}_{-0.09}$ & $29.16^{+0.28}_{-0.37}$ & $0.90^{+0.05}_{-0.03}$ & $39.88^{+0.10}_{-0.09}$ & $29.49^{+0.37}_{-0.31}$ & $1.14^{+0.07}_{-0.05}$ & $< 38.98$ & $30.48^{+0.11}_{-0.08}$ & $1.11^{+0.07}_{-0.07}$ \\[2pt]
        4${\times}$4\,kpc$^2$ & $39.87^{+0.07}_{-0.07}$ & $28.87^{+0.41}_{-0.39}$ & $0.92^{+0.04}_{-0.04}$ & $39.77^{+0.09}_{-0.11}$ & $29.29^{+0.41}_{-0.42}$ & $1.19^{+0.05}_{-0.06}$ & $< 39.45$ & $30.25^{+0.09}_{-0.09}$ & $1.04^{+0.09}_{-0.05}$ \\[2pt]
	\hline
    \end{tabular}
    \begin{tablenotes}
\small
\item NOTE: Model $\rm{log} \, \textit{L}_X = \rm{log}(10^{\alpha +
    log \, \rm{SFR}} + 10^{\beta + log \, M_\star}) + \sigma$, where
  $\alpha$ and $\beta$ are the scaling factors of the X-ray
  luminosity resulting from the young and the old stellar populations
  and $\sigma$ is a again a Gaussian random variable account for any
  intrinsic scatter in the data.
    \end{tablenotes}
    \end{threeparttable}
\end{table*}

\section{Discussion}
\label{sec:Discussion}

\subsection{Comparisons between different SFR indicators}
\label{sec:Discussion_SFR_indicators}

There is growing evidence that the X-ray emission of XRB populations
evolves as a function of time
\citep[e.g.,][]{2013ApJ...764...41F,2019ApJ...887...20A,2019ApJS..243....3L}.
HMXBs in particular are a short-lived population, and therefore their
abundance depends on SFH\null. Several recent studies have started to
explore the sensitivity of SFR inferred from different SFHs. For
example H$\alpha$ traces $\sim$10\,Myr stellar populations whereas
8\,$\mu$m and 24\,$\mu$m trace ${\gtrsim}$200\,Myr stellar
populations. However, what is not clear yet is how the X-ray scaling
relations depend on the SFH of the population responsible for the
X-ray emission, because previous works have used indiscriminately
different SFR indicators even for different galaxies in the same
scaling relations. Such variation may contribute to the observed
scatter.

\begin{figure}[h]
\begin{center}
\includegraphics[width=\columnwidth]{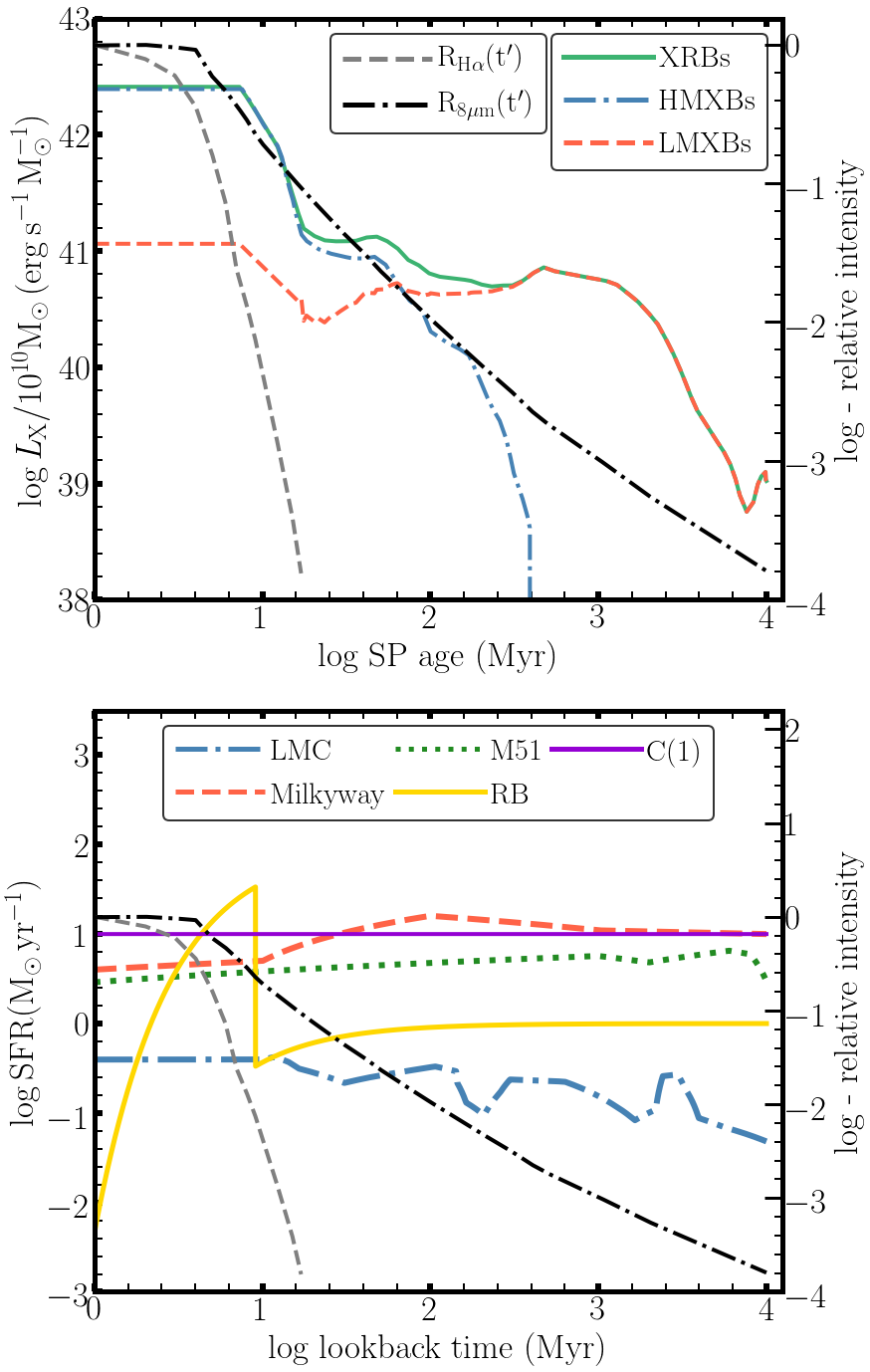}
\caption{Upper panel: Bolometric X-ray luminosity per
$M_\star$ (in units of $10^{10}$\,\msun, green line) of a stellar
population as a function of the population's age from
\citet{2013ApJ...764...41F}.  Contributions of
HMXBs are shown by the blue dashed-dotted line and of LMXBs by the red dashed line. Response functions for H$\alpha$
and 8\,$\mu$m are shown with grey dashed and black dashed-dotted
lines respectively, and their scales are shown on the right ordinate.
The 24\,$\mu$m response function is indistinguishable from the
8\,$\mu$m one. Bottom panel: Measured SFR as a function of
lookback time  for five indicative SFHs. The SFHs comprise
one representing an early-type spiral galaxy, for which we used the Milky
Way's (MW) SFH \citep{2018ApJS..237...33X}, the SFH of the Large
Magellanic Cloud (LMC) as a proxy for a dwarf galaxy dominated by a
recent star-formation episode \citep{2009AJ....138.1243H}, the SFH of
M51 \citep{2017ApJ...851...10E} as a galaxy with a peak of
star formation around 200\,Myr ago, the SFH of a galaxy with a resent
star-formation burst (RB), formulated as a double exponential model
\citep{2019A&A...622A.103B} with $t_0 = 4000$\,Myr, $t_1 =
3000$\,Myr, $\tau_0 = 1000$\,Myr, $\tau_1 = 1000$\,Myr, and
$\kappa=10$, and a galaxy with constant SFR throughout its history
with $\SFR=10^{1}$\,\msunpyr (labeled ``C(1)'') 
for reference. These SFHs are presented with red dashed, blue
dashed-dotted, green dotted, yellow, and purple lines
respectively. 
Gray dashed and black dash-dotted lines show the response functions
from the upper panel.}
    \label{fig:SFH}
\end{center}
\end{figure}

Our observations show a systematic difference in the $L_X$--SFR
correlations between the different SFR indicators. The H$\alpha$ SFR
indicator gives a steeper, more linear slope and the lowest scatter,
indicating that it is better correlated with the XRBs' X-ray emission
than the 8\,$\mu$m and the 24\,$\mu$m indicators. The H$\alpha$
emission traces the ionizing radiation from stellar populations with
ages \citep[e.g.,][]{2012ARA&A..50..531K, 2014A&A...571A..72B,
2016A&A...589A.108C} similar to the formation timescale of the HMXBs
\citep[e.g.,][]{1991PhR...203....1B, 2006csxs.book..623T,
2013ApJ...764...41F}. In contrast, the 8 and 24\,$\mu$m bands'
connection with HMXBs is diluted (Fig.~\ref{fig:SFH}) by the much
larger age range those SFR indicators reflect.

The X-ray emission from LMXBs begins to dominate over that from HMXBs
for stellar populations older than ${\gtrsim}$80\,Myr
(Fig.~\ref{fig:SFH}), even though the bulk of their population forms
at much later times. In regions dominated by a young stellar
population, the IR indicators will be dominated by the same young
stellar populations traced by the H$\alpha$ emission, which also host
the HMXB populations. On the other hand, for regions with
star-forming activity extending beyond 100 Myr, the IR indicators
will include contribution from older stellar populations than those
traced by the H$\alpha$ emission. These older stellar populations do
not include HMXBs \citep[e.g.,][]{2013ApJ...764...41F}, resulting in
increasing scatter.

In order to obtain at least a qualitative picture of the X-ray
luminosity scaling relations' dependence on SFH, we performed a simple
simulation study where we calculated the X-ray luminosity, SFR, and
stellar mass under different assumptions for the SFH\null. 
The top panel of Fig.~\ref{fig:SFH} presents the X-ray
output of a stellar population from the model of
\citet{2013ApJ...764...41F} as a function of age along with the
age sensitivities (response functions) of the three SFR indicators
considered here.\footnote{The response functions were calculated by
modeling the evolution of the H$\alpha$, 8\,$\mu$m, and 24\,$\mu$m
emission for an instantaneous burst of star formation. In order to
subtract the stellar continuum from the 8\,$\mu$m emission, we also
calculated the ratio of the flux in the 3.6\,$\mu$m and 8\,$\mu$m
\spitzer-IRAC bands for the same decaying population without
including any dust contribution. These calculations were performed
with \textit{CIGALE} v.2018.0.1 \citep{2019A&A...622A.103B}. The
stellar populations were modeled after the BC03
\citep{2003MNRAS.344.1000B} models assuming solar metallicity. We
considered models with Salpeter \citep{1955ApJ...121..161S} or
Chabrier \citep{2003PASP..115..763C} IMFs,  values for the
absorption $E(B-V)=0.3, 1.0$, nebular component ionization parameter
$U=-1.0,-2.0,-3.0,-4.0$, and two
dust emission models: those of \citealt{2014ApJ...784...83D} and
\citealt{2007ApJ...657..810D}. We explored
different values of the $\alpha$ parameter in the
\citep{2014ApJ...784...83D} dust model and of the PAH mass fraction
(\textit{qpah}) and limiting ionization field ($U_{\rm{min}}$) for
the \cite{2014ApJ...780..172D} models. The response functions presented in
Fig.~\ref{fig:SFH} are the average of the results from the different
models. A more detailed discussion of the response functions and the
parameters they depend on will be presented in Kouroumpatzakis \etal\ (in
prep). Similar investigations for various SFR indicators have been
presented in previous works
\citep[e.g.,][]{2016A&A...589A.108C,2014A&A...571A..72B} but for
different SFR indicators than those used here or for more complex
SFHs, which complicate the disentanglement of the contribution of
different stellar populations to the measured SFR\null.}

\begin{figure}
\begin{center}
\includegraphics[width=\columnwidth]{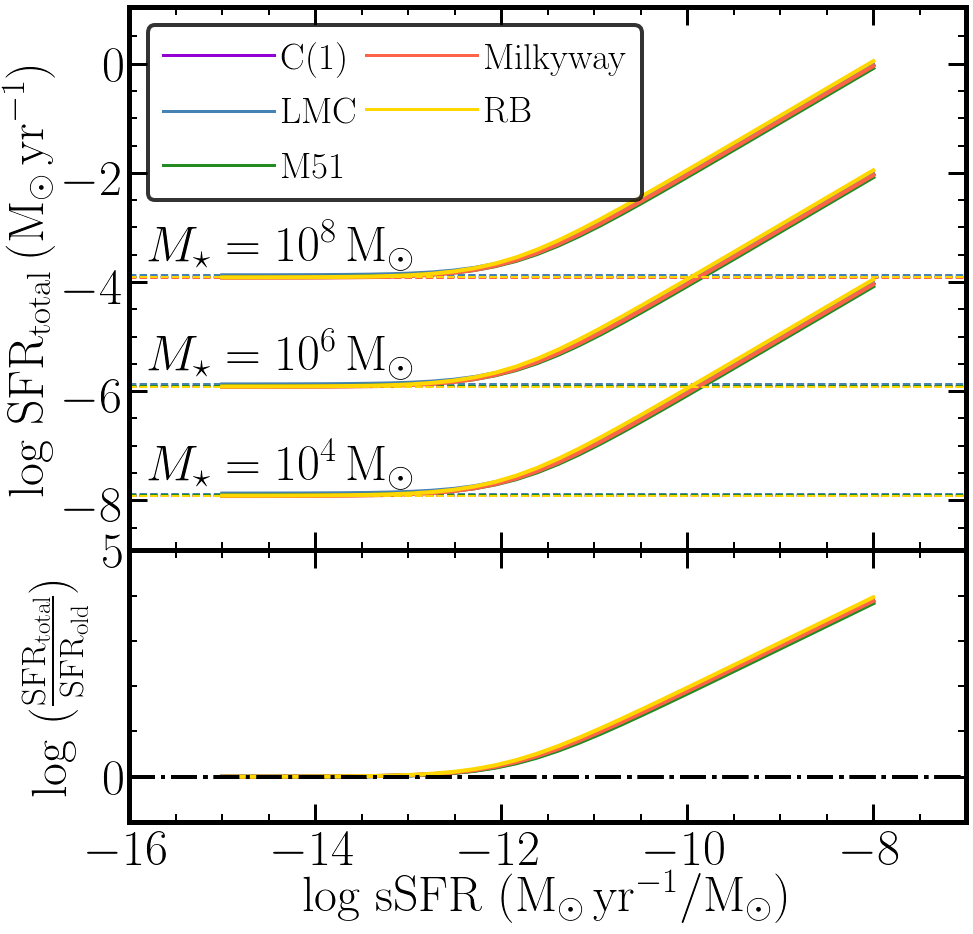}
\caption{Upper panel: Inferred H$\alpha$-based SFR separating the
    contribution of young and old stellar populations as a function
    of sSFR for three stellar population masses. The lines are based
    on \textit{CIGALE} simulations using the five SFHs presented in
    Fig.~\ref{fig:SFH} with {young} and {old} stellar populations
    separated at 100\,Myr.  Solid lines represent total SFR, and
    dashed lines represent the contribution of the old stellar
    populations.  Results are nearly identical for all SFHs. Bottom panel:
    ratio  $\rm{SFR_\textit{total}/SFR_\textit{old}}$ as a
    function of sSFR\null.}
    \label{fig:SFR_sSFR_m}
\end{center}
\end{figure}

\begin{table*} \centering
	\caption{Results of the calculations based on the model of
\citealt{2013ApJ...764...41F}, the H$\alpha$, 8\,$\mu$m, 24\,$\mu$m
response functions, and the SFHs shown in Figure \ref{fig:SFH}.}
	\label{tab:SFH_table} 
	\begin{tabular}{lcccccc} & & MW & M51 & $C(1)$ & LMC & RB\\
    \hline $\rm{\log} \, \textit{M}_\star (M_\odot)$ & Eq.~\ref{eq:Mstarx} & 10.74
    & 11.02 & 11.00 & 9.03 & 10.44\\ $\rm{SFR_{H\alpha} \, (M_\odot
    yr^{-1})}$ & Eq.~\ref{eq:SFRx} & 3.17 & 4.30 & 10.00 & 0.40 & 6.09\\
    $\rm{SFR_{8 \, \mu m} \, (M_\odot yr^{-1})}$ & & 3.34 & 4.72 & 10.00
    & 0.39 & 13.24\\ $\rm{SFR_{24 \, \mu m} \, (M_\odot yr^{-1})}$ & &
    3.35 & 4.74 & 10.00 & 0.39 & 13.53\\ $\rm{t_{eff} \, (H\alpha; \,
    Myr)}$ & Eq.~\ref{eq:Teff} & 10 & 9 & 9 & 9 & 7\\ $\rm{t_{eff} \, (8
    \, \mu m; \, Myr)}$ & & 709 & 569 & 594 & 272 & 194\\ $\rm{t_{eff} \,
    (24 \, \mu m; \, Myr)}$ & & 716 & 574 & 601 & 276 & 204\\
    log\,$L_{X(\rm{XRBs})} \, \rm{(erg \, s^{-1})}$ & Eq.~\ref{eq:LXx} &
    42.99 & 43.15 & 43.45 & 42.03 & 43.73\\ log\,$L_{X(\rm{HMXBs})} \,
    \rm{(erg \, s^{-1})}$ & & 42.95 & 43.09 & 43.42 & 42.01 & 43.72\\
    log\,$L_{X(\rm{LMXBs})} \, \rm{(erg \, s^{-1})}$ & & 41.97 & 42.27 &
    42.35 & 40.85 & 42.43\\ $\rm{\log \,\alpha' / \beta' \, (H\alpha; \,
    M_\odot \, yr^{-1} / M_\odot})$ & Eq.~\ref{eq:ab} &
    $-10.25$ & $-10.40$ & $-10.01$ & $-9.44$ & $-9.66$\\ $\rm{\log
    \,\alpha' / \beta' \, (8 \, \mu m; \, M_\odot \, yr^{-1} /
    M_\odot})$ & & $-10.23$ & $-10.36$ & $-10.01$ & $-9.45$ &
    $-9.33$\\ $\rm{\log \,\alpha' / \beta' \, (24 \, \mu m; \, \,
    M_\odot \, yr^{-1} / M_\odot})$ & & $-10.23$ &
    $-10.36$ & $-10.01$ & $-9.45$ & $-9.32$\\ \hline
	\end{tabular}
\end{table*}

Based on the XRB luminosity evolution and the SFR indicator response
functions, we can quantify the dependence of the $L_X$--SFR relations
on the SFH and the SFR indicator used. To demonstrate this effect we
considered five different SFHs (see Fig.~\ref{fig:SFH}). The total
stellar mass is:
\begin{eqnarray}
   M_\star = \int_{0}^{t} {\SFH}({t}') dt' \quad,
  \label{eq:Mstarx}
\end{eqnarray}
and the ``effective'' SFR  for each indicator, which accounts for their
sensitivity to older or younger stellar populations is: 
\begin{eqnarray}
   {\rm SFR_\chi} = \frac{\int_{0}^{{t}} {\SFH}(t') R_\chi (t')
     \,dt'}{\int_{0}^{t} R_\chi (t')  \,dt'} \quad . 
\label{eq:SFRx}
\end{eqnarray}
Figure~\ref{fig:SFH} shows five example SFHs, and results for each one
are presented in Table~\ref{tab:SFH_table}. We expect variations in
the SFR for the different SFH scenarios only if the SFR changes
within the time window of each indicator (e.g., largest difference
for the RB example).

Another way to show differences in the average stellar population
ages traced by the different SFR indicators $\chi$ is the effective
age of the stellar population for given SFH:
\begin{eqnarray}
t_{\rm{eff,\chi}} = \frac{\int_{0}^{t}  t' R_\chi (t')  \SFH({t}') \,  dt'}
{\int_{0}^{t} R_\chi(t') \SFH(t') \, dt'} \quad .
\label{eq:Teff}
\end{eqnarray}
H$\alpha$ emission traces the youngest stellar populations
($t_{\rm{eff}}\le 10$\,Myr: Table~\ref{tab:SFH_table}) almost
unaffected by the different SFHs. When there is a recent burst of
star formation, the IR indicators trace stellar populations with
younger average ages (e.g., for RB $t_{\rm{eff}}\simeq200$\,Myr),
but when the SFH is not dominated by a recent star-formation burst,
the same indicators trace much older stellar populations
($t_{\rm{eff}}\simeq600$\,Myr).

Although the 8 and 24\,$\mu$m SFR-indicator response
functions trace fairly well the HMXB X-ray luminosity as a function
of time (Fig.~\ref{fig:SFH}), they can be affected by emission from
stars older than those that can form HMXBs. Thus
these  indicators can overestimate the SFR when a stellar
population is dominated by older stars and has larger
$t_{\rm{eff}}$ (Table~\ref{tab:SFH_table}).  In addition, because
${\gtrsim}$60\,Myr populations do not contribute to the
formation of HMXBs
\citep[e.g.,][]{2013ApJ...764...41F,2018MNRAS.479.3526G,2019arXiv190101237A},
the $L_X$--SFR scaling relations based on the 8 and 24\,$\mu$m
indicators will result in lower scaling factors for galaxies with
SFHs not dominated by a recent burst. Therefore, H$\alpha$
is the most appropriate proxy to trace the young HXMB populations as
demonstrated by the tighter H$\alpha$-based scaling relations
(Table~\ref{tab:Fit_results}).

All of the SFR indicators can break down in regions
with extremely low SFR\null. In such regions, UV photons
originating from A-type or post-AGB stars may give significant
contributions.  The UV luminosity  emitted by a stellar population is
the sum of
the emission from \textit{young} and \textit{old} stars.
The H$\alpha$ SFR indicator is based on the {\em number} of
Lyman continuum photons, assuming that all the Lyman photons are
absorbed by the gas \citep[case-B
recombination;][]{2006agna.book.....O}.  The 8\,\micron\ indicator
is based on the number of photons at somewhat longer UV
wavelengths, while the 24~\micron\ indicator is based on the UV
{\em luminosity}, assuming all the energy is absorbed by dust and
reradiated.

In order to quantify the contribution of older stellar
populations when measuring extremely low SFRs from H$\alpha$, we calculated
separately the
SFRs that would be measured for the old and young populations in the
aforementioned \textit{CIGALE} simulations for the five
SFH scenarios. We considered as  \textit{young} stars
with ages $<$100~Myr and  the rest as \textit{old}.
Lyman-continuum photons produced by each population were converted to
the equivalent SFR via the \cite{1998ARA&A..36..189K} factor.
Dividing by stellar mass gave the equivalent sSFR\null.
The results are shown in Figure~\ref{fig:SFR_sSFR_m}. Older stellar
populations make no significant contribution to the ionizing-photon
budget in regions with
$\textrm{sSFR}\gtrsim10^{-12}$\,\msunpyr/\msun.  Even this upper
limit assumes that all UV photons from the older stellar populations
contribute to the ionization of the interstellar medium, but in real
spiral galaxies, many such photons escape. Therefore, the derived
limiting sSFR is a conservative limit for trustworthy SFRs from young
stellar populations, but lower sSFR than this value cannot be
reliably measured by H$\alpha$. 
This limiting sSFR is insensitive to the SFH. 
The corresponding limiting SFR of course depends on stellar mass. 
At the sSFRs  of the most actively star-forming regions in our sample the ionizing photon production rate exceeds that of the old by 4dex.
For the present study, as shown in Fig.~\ref{fig:sfr_ssfr_range}, at most 3.5\% of the 
regions (and fewer for the regions smaller than 4${\times}$4\,kpc$^2$)
have $\textrm{sSFR}{<}10^{-12}$\,\msunpyr/\msun, indicating that UV
photons from older stellar populations do not affect our present
conclusions.

The X-ray luminosity for each SFH
scenario (Fig.~\ref{fig:SFH}) is:
\begin{eqnarray} 
L_{X}^{\upsilon}(t) = 
\int_{0}^{t} {\bigg(\frac{L_X(t')}{M_\star}\bigg)_\upsilon}
{\SFH}(t')\,dt' \quad,
\label{eq:LXx}
\end{eqnarray}
where $\upsilon$ indicates the particular XRB
population (HMXBs, LMXBs, XRBs), and $M_\star$ is the total stellar
mass of the parent stellar population of the XRBs.  The results of
these calculations show ${\ge}$0.85\,dex differences in the X-ray
luminosity produced by the HMXBs and LMXBs regardless of the SFH
assumed. This difference is larger for SFHs with more intense and
more recent star-formation episodes.

A metric of the relative contribution of HMXB and LMXB populations in
the integrated X-ray luminosity is the ratio ($\alpha/\beta$) used in
Eq.~\ref{Eq:mdl3}. Given that $L_{X,\rm{HMXB}} = \alpha\SFR$, and
$L_{X,\rm{LMXB}} = \beta  M_\star$, we can calculate the
theoretically expected $\alpha'/\beta'$ ratio from the X-ray
luminosity of the LMXB and HMXB populations given an SFH
(Eq.~\ref{eq:LXx}).
\begin{eqnarray}
(\alpha'/\beta')_\chi =
\frac{L_{X,\rm{HMXB}}}{\rm{SFR_\chi}} \bigg/
\frac{L_{X,\rm{LMXB}}}{M_\star}
    \label{eq:ab}
\end{eqnarray}
for each SFR indicator (Eq.~\ref{eq:SFRx}). The results for these
calculations are presented in Table~\ref{tab:SFH_table}. The
continuous SFH gives 
$\alpha'/\beta' = 10^{-10.01}$\,\msunpyr/\msun. 
LMC-like or
RB-like SFHs, with a recent star-formation episode, show
$\alpha'/\beta' > 10^{-10.01}$\,\msunpyr/\msun. 
On the other hand, MW and M51, which comprise far
older stellar populations, show
$\alpha'/\beta' < 10^{-10.01}$\,\msunpyr/\msun, indicating a
larger contribution of LMXBs to the total X-ray luminosity.

\subsection{Distributions of X-ray luminosity for regions with
different sSFR}
\label{sec:LX_excess}

\begin{figure}
    \begin{center}
\includegraphics[width=\columnwidth]{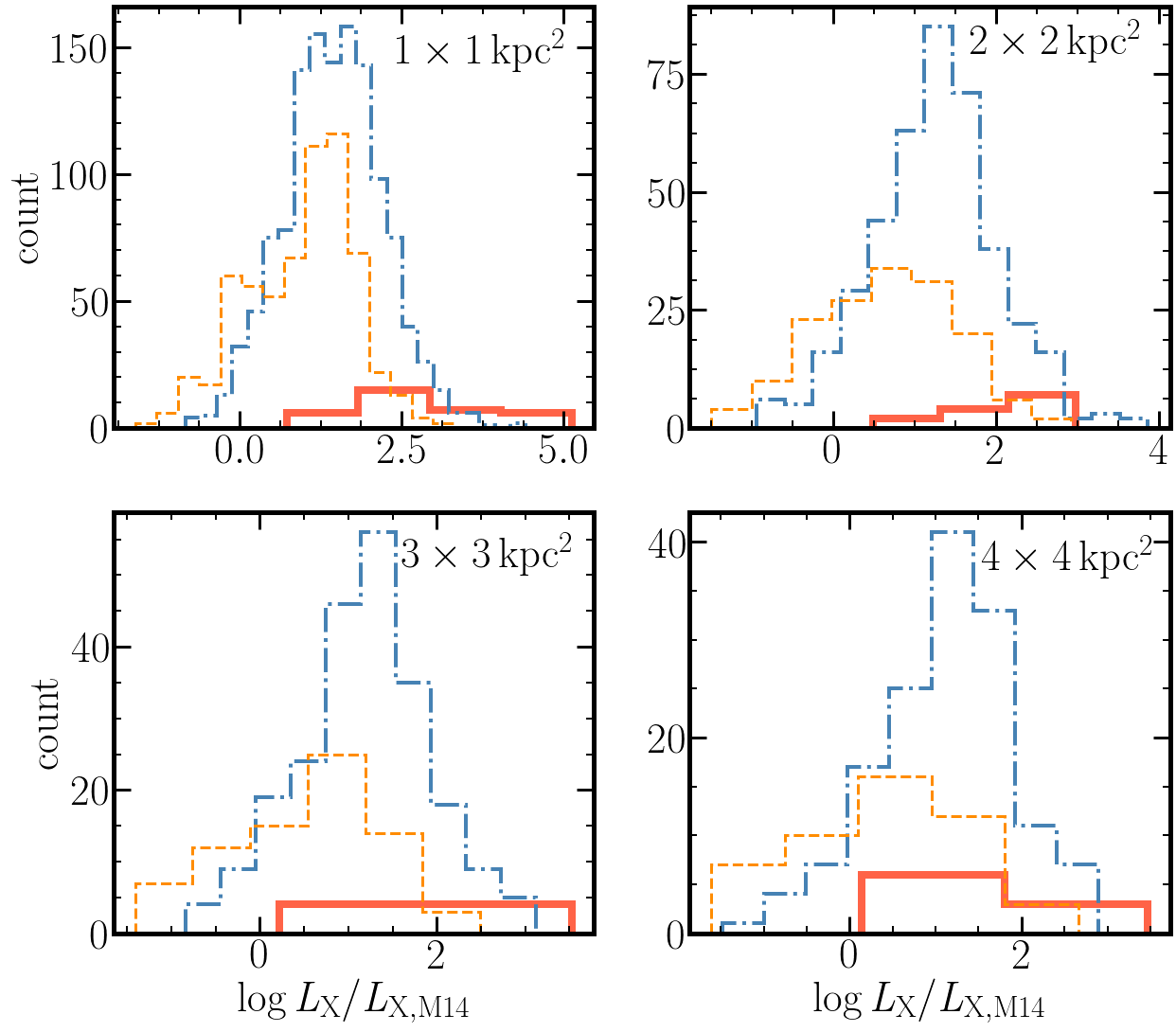}
\caption{Histograms of excess X-ray luminosity relative to the M14
relation (Eq.~\ref{eq:L_excess}). Panels show the distributions for
sub-galactic regions as labeled. Regions with $\textrm{sSFR} \ge
10^{-10}$, $\rm{10^{-12} \le sSFR \le 10^{-10}}$, and $\rm sSFR
\le 10^{-12}$ are represented by orange dashed, blue dashed-dotted,
and thick red lines respectively. The H$\alpha$-based SFR was used
here.}
    \label{fig:Hist_Lx_Exceess}
    \end{center}
\end{figure}

If the X-ray emission arises from a population of HMXBs, it would be
expected to scale linearly with SFR\null. The scaling factor depends on
the formation efficiency of HMXBs and their integrated luminosity per
unit SFR, which is a function of their age (Figure \ref{fig:SFH},
Section \ref{sec:Discussion_SFR_indicators}). Therefore, the
galaxy-wide scaling relations are expected to extend to lower SFR
even on sub-galactic scales if the average properties of the stellar
populations (age and metallicity) are the same. Any deviations from
this linear relation or change in slope indicates a different XRB
population. As discussed in Section \ref{sec:LX_SFR_Correlations}, we
observe an excess of X-ray emission in the low SFR regime compared to
the extrapolation of the linear $L_X$--SFR relation from higher
SFR\null. The excess can be quantified as the ratio of the measured
luminosity to the one expected from the linear scaling relation of
M14,
\begin{eqnarray}
L_{X, \rm{excess}} = \log  L_X / L_{X, \rm{M14(SFR)}} \quad .
	\label{eq:L_excess}
\end{eqnarray}
Fig.~\ref{fig:Hist_Lx_Exceess} shows histograms of the
excess in regions of different sSFR\null. The modes and 68.3\% confidence
intervals of these distributions are presented in 
Table~\ref{tab:Excess_table}. Regions with lower sSFR exhibit
systematically higher excess, including the highest values seen. The
bin of ${\rm sSFR} \le 10^{-12}$, in particular,
isolates sub-galactic regions with very low current star formation,
where no massive young stars and consequently HMXBs are expected. At
these sSFRs, the dominant source of X-ray emission is expected to be
LMXBs \citep[e.g.,][]{2010ApJ...723..530P}.

\begin{figure}
\begin{center}
\includegraphics[width=0.8\columnwidth]{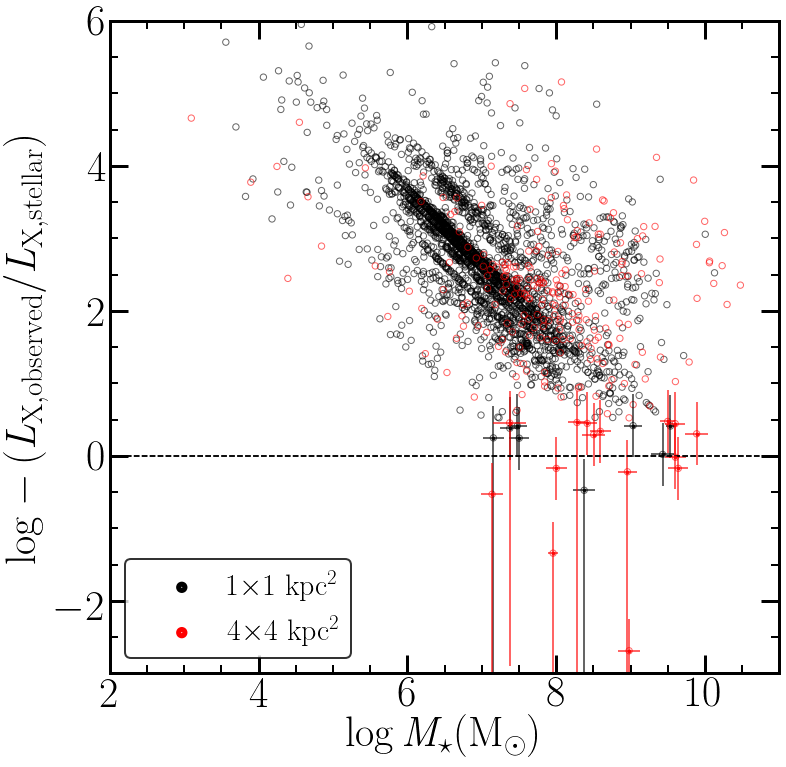}
\caption{Observed X-ray luminosity over the expected stellar X-ray
luminosity (Eq.~\ref{eq:L_boronson}) as a function of the stellar
mass for sub-galactic regions of $1 \, {\times} \, 1$ (black) and
$\rm{4 \, {\times} \, 4 \, {kpc}^2}$ (red). Error bars are shown for
regions with
$\log(L_{X,\rm observed}/L_{X,\rm stellar}) \le
0.5$. Other regions are represented only by circles to avoid
clutter.}
    \label{fig:Stellar_Lx}
    \end{center}
\end{figure}

In regions encompassing
large enough stellar mass, the collective emission of cataclysmic
variables (CVs) and coronally active binaries (ABs) may have
non-negligible contribution, particularly at the very low integrated
X-ray luminosities probed (${\le}10^{35.5}$\,\erg). The relation between the X-ray luminosity from these
components ($L_{X,\rm{stellar}}$) and $K$-band luminosity
\citep{2011ApJ...729...12B} is:
\begin{eqnarray}
\frac {L_{X, \rm{stellar}}}{(\erg)} =
9.5^{+2.1}_{-1.1} \times 10^{27}  {L}_{{K}\odot}
	\label{eq:L_boronson}
\end{eqnarray}
where $L_{K\odot}$ is in solar luminosities (a proxy
of the total stellar mass they encompass). Because in this work we
used 3.6\,$\mu$m as a proxy of stellar mass, we converted 
3.6$\,\mu$m to \textit{K}-band luminosities.\footnote{The 3.6\,$\mu$m to
\textit{K}-band magnitudes were calibrated and converted using the
complete SFRS. The linear correlation found is:
\begin{eqnarray} m_{K} = 1.876 \pm 0.1 + 1.10 \pm 0.01  m_{3.6
\micron}\quad.
  \end{eqnarray}} For most of the regions, CVs' and ABs' stellar
contribution to the X-ray luminosity is less than observed by more
than 1~dex (98\%, 95\%, 91\%, and 90\% of  the
1${\times}$1, 2${\times}$2, 3${\times}$3, and 4${\times}$4~kpc$^2$
regions respectively), even for regions with extremely high stellar 
mass (Fig~\ref{fig:Stellar_Lx}). However, there are a handful of
regions where the calculated stellar X-ray luminosity is comparable
to the observed X-ray luminosity, but they also exhibit high
relative uncertainties. This minority of regions is not sufficient to
explain the observed X-ray luminosity excess. Alternatives being
insufficient, the bulk of the X-ray luminosity excess found in the
low SFR regime must come from LMXB emission.

\begin{table} \centering
\caption{Median excess $L_{X,0.5-8\rm{keV}}$ over the expected by M14
in bins of different sSFR\null. The number of sub-galactic regions
included in each bin is given in parentheses.}
	\label{tab:Excess_table} 
\setlength\tabcolsep{2.5pt} 
	\begin{tabular}{lcccc} 
\hline\hline
Size\,(kpc$^2$) & $\rm{sSFR {\le}
10^{-12}}$ & $\rm{10^{-12} {\le} sSFR {\le} 10^{-10}}$ &
$\rm{10^{-10} {\le} sSFR}$\\ 
\hline $1 \times 1$ & $2.73\pm1.08$ (34)
& $1.45 \pm0.77$ (1263) & $1.1 \pm0.83$ (617)\\ $2 \times 2$ & $2.24
\pm0.95$ (14) & $1.25 \pm0.77$ (403) & $0.71 \pm0.84$ (157)\\ 
$3 \times 3$ & $2.06 \pm1.1$ (8) & $1.23 \pm0.75$ (225) & $0.68 \pm0.89$
(76)\\ 
$4 \times 4$ & $1.6 \pm1.08$ (9) & $1.15 \pm0.78$ (146) &
$0.56 \pm0.94$ (48) \\
\hline
\end{tabular}
\end{table}

\subsection{Comparison with galaxy-wide scaling relations}
\label{sec:Discussion_integrated_emission}
 
Sub-galactic regions show a shallower slope of $L_X$--SFR (Table
\ref{tab:Fit_results}, Fig.~\ref{fig:Fit_Lx_SFR_all}) compared to the
M14 relation for all cases considered in this work. This is driven by
regions with high X-ray luminosity at SFR\,$\rm{{\le} 10^{-3} \,
M_{\odot} yr^{-1}}$, particularly at the smallest physical
scales. For reference, the lowest SFR used in the derivation of the
galaxy-wide scaling relation was ${\sim}10^{-1}$\,\msunpyr,
whereas our analysis extends to 5\,dex
lower SFR\null. The X-ray emission of these regions arises from an
unresolved population of LMXBs (Section \ref{sec:LX_excess}). The
inclusion of the stellar mass as a parameter (Eq.~\ref{Eq:mdl3})
accounts for the LMXB contribution, particularly in regions with low
SFR or those dominated by older stellar populations (low sSFR). As a
result we obtain good fits with linear scaling of the X-ray
luminosity with respect to both the SFR and stellar mass.

Even though our $L_X$--SFR--$M_\star$ fits follow a different
approach from L16, by fitting sub-galactic regions and including an
intrinsic scatter term (Eq.~\ref{Eq:mdl3}), our results are in good
agreement (Fig.~\ref{fig:LxSFR_sSFR}) with only small differences in
the best-fit parameters. The main difference is that we find
significant intrinsic scatter. We interpret the scatter as the result
of stochastic effects. In all cases, the H$\alpha$ SFR indicator
gives the lowest scatter and the best agreement with the relation of
L16 (despite their use of UV and far-IR instead of H$\alpha$-based
SFR tracers). However, we do find differences with the
\citealt{2019ApJS..243....3L} scaling relations, which are based on
integration of the XRB luminosity functions (XLFs) derived for
different sSFR regimes. More specifically, while for the largest
physical scales (4${\times}$4\,kpc$^2$) and the scaling with SFR
(parameter $\alpha$) in the $L_X$--SFR--$M_\star$ fit
(Eq.~\ref{Eq:mdl3}) we find good agreement for all SFR indicators
used, in the case of smaller physical scales, we find increasing
$L_X$--SFR scaling factors (Table~\ref{tab:Fit_results_2D}).
This can be explained by the local variations of stellar populations
between the different regions, which results in localised variations
of the $L_X$/SFR scale factor (e.g., Section
\ref{sec:Discussion_SFR_indicators}). This effect in combination with
stochastic sampling of the XLF results in a few regions with high
X-ray luminosity (because of the presence of very young populations
and/or luminous individual sources) and therefore small $L_X$ and SFR
uncertainties, that can drive the fits to steeper slopes. At larger
scales, local variations in the X-ray emission and stellar
populations are averaged out, and the scaling relations approach the
galaxy-wide relations.  On the other hand, the $L_X$--$M_\star$
scaling (parameter $\beta$ in Eq.~\ref{Eq:mdl3}) is consistent with
\citet{2019ApJS..243....3L} for most SFR indicators and spatial
scales we consider.  The smoother spatial distribution of the older
stellar populations and the weak $L_X$--age dependence of the X-ray
binaries associated with them results in more uniform sampling
regardless of physical scales and therefore consistent $L_X$--$M_\star$
scaling factors through the different physical scales.

\subsection{Intrinsic scatter \& stochasticity}
\label{sec:Discussion_intrinsic_scatter}

The wide range of SFRs and stellar masses probed in our study
(Fig.~\ref{fig:sfr_ssfr_range}, Table~\ref{tab:Sample}) is ideal for
examining the intrinsic scatter under conditions found in nearby
galaxies. This scatter could be the result of
(a) Poisson sampling of sparsely populated luminosity functions or
(b) time variability of XRBs \cite[e.g.,][]{2004MNRAS.349..146G}.
Such scatter has been previously reported in galaxy-wide scaling relations,
particularly at lower SFRs
\citep[e.g.,][]{2014MNRAS.437.1698M,2019ApJS..243....3L}. However, as
discussed in Section \ref{sec:Discussion_SFR_indicators}, an
additional source of scatter could be stellar population differences
through their effect on the inferred SFR and the age-dependent X-ray
output of stellar populations.

There is intrinsic scatter in the sub-galactic $L_X$--SFR 
(Table~\ref{tab:Fit_results}) and $L_X$--SFR--$M_\star$ 
(Table~\ref{tab:Fit_results_2D}) correlations. However, we do not find any
evidence for anti-correlation of the intrinsic scatter with the SFR
(Table~\ref{tab:Fit_results_2}) as would be expected from stochasticity
or time variability. This could be the result of the large uncertainties
in the SFR and X-ray luminosity measurements for the individual
regions at low SFR, which could mask any such trend. On the other
hand, the overall intrinsic scatter we measure both in the $L_X$--SFR
and the $L_X$--SFR--$M_\star$ relations (typically 0.5--1.0 dex) is
larger than the scatter observed in the galaxy-wide relations (e.g.,
${\le}$0.37 dex in L16). This additional scatter could be the result
of bright X-ray sources in some of the individual regions. However,
typically less than 3\% of the regions in each galaxy of our sample
encompass individually detected X-ray sources, making them an
unlikely source for the increased scatter on sub-galactic regions.

One parameter that is particularly important on sub-galactic scales
is local variations of the stellar populations, such as those
resulting from the spiral structure, localized star-formation
episodes, sequential star formation, and metallicity gradients. XRB
population synthesis models show that the X-ray emission for an
ensemble of XRBs is a strong function of the age and metallicity of
their parent stellar populations
\citep[e.g.,][]{2013ApJ...764...41F,2006MNRAS.370.2079D,2010ApJ...725.1984L,2019ApJS..243....3L}. This
is supported by observational studies of the XRB populations
associated with different stellar generations
\citep[e.g.,][]{2016MNRAS.459..528A,2019ApJ...887...20A} or
populations of different metallicity
\citep[e.g.,][]{2010MNRAS.408..234M,2013ApJ...769...92P,2015A&A...579A..44D,2016MNRAS.457.4081B}. On
galaxy-wide scales, any local variations of the stellar populations
and the corresponding X-ray emission can be smeared out giving an
average $L_X$/SFR value for the entire galaxy. On the other hand,
local variations of the stellar populations within a galaxy (which
can vary in age from a few Myr for very young star forming regions to
several Gyr for interarm regions) can result in very different X-ray
emission efficiency as discussed in Section
\ref{sec:Discussion_SFR_indicators}.

An additional source of scatter could be local variations of
absorption. In order to correct for this one would need spatially
resolved extinction and $N_{\rm{H}}$ maps from X-ray spectral fits in
each sub-galactic region, which are not available for these data
(c.f. Section \ref{sec:X-rays}). Furthermore as discussed in Section
\ref{sec:LX_SFR_Correlations}, the absorption in H$\alpha$ and soft
X-rays is similar, which reduces the effect of differential
extinction across the galaxies.

A general trend is that scaling relations based on the H$\alpha$
emission show lower scatter than the relations based on the 8\,$\mu$m
and 24\,$\mu$m SFR indicators. H$\alpha$ emission traces the stellar
populations with ages ${\sim}$10\,Myr (Fig.~\ref{fig:SFH}; 
Table~\ref{tab:SFH_table}) which are most relevant to the HMXBs (which have
lifetimes ${\le} 30 \, \rm{Myr}$; Section
\ref{sec:LX_SFR_Correlations}). On the other hand, although
the IR-based SFR indicators still trace young stellar populations,
they are sensitive to a much wider range of ages. Therefore, they are not a clean proxy for the
star-formation episodes that produced the HMXBs. This mismatch
between the formation timescales of the HMXBs and the star-formation
timescales probed by the different SFR indicators could be the origin
of the larger scatter we measure in the sub-galactic scaling
relations in comparison to the galaxy-wide relations. This is because
sub-galactic regions may have significant variations in their SFHs
compared to the overall galaxy averages.

\section{Summary}
\label{sec:Summary}

We present scaling relations between $L_X$--SFR--$M_\star$ on
sub-galactic scales using a maximum likelihood method that takes into
account the posterior (not necessarily Gaussian) uncertainty
distributions of all the data. In this way we obtain unbiased scaling
relations by including in our analysis regions that have extremely
low SFRs, stellar masses and X-ray luminosities which otherwise would
be omitted. This analysis extends the $L_X$--SFR and the
$L_X$--SFR--$M_\star$ relations down to 
$\rm SFRs\simeq 10^{-6}$\,\msunpyr,
and $\rm sSFRs \simeq 10^{-14}$\,\msunpyr/\msun. These are 5\,dex and 
2\,dex lower in SFR and sSFR respectively than existing galaxy-wide
scaling relations. In the $L_X$--SFR correlation, slopes are shallower
than linear on all sub-galactic scales (1${\times}$1, 2${\times}$2,
3${\times}$3, and 4${\times}$4\,kpc$^2$) and by all SFR indicators
(H$\alpha$, 8\,$\mu$m, and 24\,$\mu$m) used in this work. This
shallower slope is driven by regions with high X-ray luminosity at
low SFR (${\le} 10^{-3}$\,\msunpyr), probably due to
a population of LMXBs. For larger sub-galactic regions, correlations
of $L_X$--SFR converge to the integrated galactic emission relations.

The full-band X-ray luminosity fits are very similar to those of the
hard band. Although the use of the full X-ray band increases the
scatter in the correlations, it integrates more flux and therefore
can be very useful for low-X-ray-luminosity objects. The extended
relations we present can be used to model the X-ray output of
extremely low-SFR galaxies. However, one should be careful about two
effects:

\noindent (a) Excess X-ray luminosity at SFRs\,${\le} 10^{-3} \,
{\rm{M_\odot \, yr ^{-1}}}$ requires accounting for LMXBs by using
the $L_X$--SFR--$M_\star$ relation. The $L_X$--SFR scaling relation
will be inaccurate because of this older population (Section
\ref{sec:LX_excess}).

\noindent (b) There is strong dependence of the SFR indicators on the
SFHs of the galaxies (e.g., Section
\ref{sec:Discussion_SFR_indicators};
\citealt{2014A&A...571A..72B}). The same holds for the X-ray output
of a stellar population as a function of its age or metallicity. This
is particularly important for dwarf galaxies that might be dominated
by star-formation bursts at different epochs. In order to mitigate
these effects when studying the connection between X-ray luminosity
and stellar populations, ideally one should use the SFH of a galaxy
instead of an instantaneous SFR metric \citep[e.g.,][]{2019ApJ...887...20A}.

We find no evidence for increasing intrinsic scatter in regions of
lower SFR, but the overall scatter of the $L_X$--SFR--$M_\star$
correlations is larger than galaxy-wide relations. We attribute this
to local variations of the SFH within a galaxy. The intrinsic scatter
measured depends on the SFR indicator and X-ray band used. The
combination of the hard band and H$\alpha$-based SFR shows the
tightest correlation and the smallest intrinsic scatter in both the
$L_X$--SFR and $L_X$--SFR--$M_\star$ correlations. For individual
galaxies at very low SFRs, stochastic sampling of the IMF, the XLF,
and source variability may result in increased scatter in their
integrated X-ray luminosity. However, the scaling relations we derive
should hold \textit{on average} for the low-SFR population (subject
to the caveats discussed above).
    
\section*{Acknowledgements}

The authors thank the anonymous referee for
comments that helped to improve the clarity
of the paper. K. K., A. Z., and K. K. acknowledge funding from the
European Research Council under the European Union's Seventh
Framework Programme (FP/2007-2013)/ERC Grant Agreement n. 617001
(A-BINGOS). This project has received funding from the European
Union's Horizon 2020 research and innovation programme under the
Marie Sklodowska-Curie RISE action, grant agreement No 691164
(ASTROSTAT). AZ also acknowledges support from \textit{Chandra} grant
G02-3111X.

K. K. thanks Dan Foreman-Mackey for insightful suggestions regarding
the \texttt{emcee} package and MCMC fits (private
communication). This work  used data obtained with the
\textit{Spitzer} Space Telescope, which is operated by the Jet
Propulsion Laboratory, California Institute of Technology under a
contract with NASA. This research has made use of the NASA/IPAC
Extragalactic Database (NED), which is operated by the Jet Propulsion
Laboratory, California Institute of Technology, under contract with
the National Aeronautics and Space Administration.





\begin{thebibliography}{}
\bibitem[\protect\citeauthoryear{{Alam} et~al.,}{{Alam}
  et~al.}{2015}]{2015ApJS..219...12A}
{Alam} S.,  et~al., 2015, \mn@doi [\apjs] {10.1088/0067-0049/219/1/12}, \href
  {http://adsabs.harvard.edu/abs/2015ApJS..219...12A} {219, 12}
\end{thebibliography}


\begin{thebibliography}{}
\makeatletter
\relax
\def\mn@urlcharsother{\let\do\@makeother \do\$\do\&\do\#\do\^\do\_\do\%\do\~}
\def\mn@doi{\begingroup\mn@urlcharsother \@ifnextchar [ {\mn@doi@}
  {\mn@doi@[]}}
\def\mn@doi@[#1]#2{\def\@tempa{#1}\ifx\@tempa\@empty \href
  {http://dx.doi.org/#2} {doi:#2}\else \href {http://dx.doi.org/#2} {#1}\fi
  \endgroup}
\def\mn@eprint#1#2{\mn@eprint@#1:#2::\@nil}
\def\mn@eprint@arXiv#1{\href {http://arxiv.org/abs/#1} {{\tt arXiv:#1}}}
\def\mn@eprint@dblp#1{\href {http://dblp.uni-trier.de/rec/bibtex/#1.xml}
  {dblp:#1}}
\def\mn@eprint@#1:#2:#3:#4\@nil{\def\@tempa {#1}\def\@tempb {#2}\def\@tempc
  {#3}\ifx \@tempc \@empty \let \@tempc \@tempb \let \@tempb \@tempa \fi \ifx
  \@tempb \@empty \def\@tempb {arXiv}\fi \@ifundefined
  {mn@eprint@\@tempb}{\@tempb:\@tempc}{\expandafter \expandafter \csname
  mn@eprint@\@tempb\endcsname \expandafter{\@tempc}}}


\bibitem[\protect\citeauthoryear{{Alam} et~al.,}{{Alam}
  et~al.}{2015}]{2015ApJS..219...12A}
{Alam} S.,  et~al., 2015, \mn@doi [ApJ] {10.1088/0067-0049/219/1/12}, \href
  {http://adsabs.harvard.edu/abs/2015ApJS..219...12A} {219, 12}

\bibitem[\protect\citeauthoryear{{Anastasopoulou}, {Zezas}, {Gkiokas}  \&
  {Kovlakas}}{{Anastasopoulou} et~al.}{2019}]{2019MNRAS.483..711A}
{Anastasopoulou} K.,  {Zezas} A.,  {Gkiokas} V.,   {Kovlakas} K.,  2019,
  \mn@doi [\mnras] {10.1093/mnras/sty3131}, \href
  {https://ui.adsabs.harvard.edu/abs/2019MNRAS.483..711A} {483, 711}

\bibitem[\protect\citeauthoryear{{Antoniou} \& {Zezas}}{{Antoniou} \&
  {Zezas}}{2016}]{2016MNRAS.459..528A}
{Antoniou} V.,  {Zezas} A.,  2016, \mn@doi [\mnras] {10.1093/mnras/stw167},
  \href {https://ui.adsabs.harvard.edu/abs/2016MNRAS.459..528A} {459, 528}

\bibitem[\protect\citeauthoryear{{Antoniou} et~al.,}{{Antoniou}
  et~al.}{2019a}]{2019arXiv190101237A}
{Antoniou} V.,  et~al., 2019a, arXiv e-prints, \href
  {https://ui.adsabs.harvard.edu/abs/2019arXiv190101237A} {p. arXiv:1901.01237}

\bibitem[\protect\citeauthoryear{{Antoniou} et~al.,}{{Antoniou}
  et~al.}{2019b}]{2019ApJ...887...20A}
{Antoniou} V.,  et~al., 2019b, \mn@doi [\apj] {10.3847/1538-4357/ab4a7a}, \href
  {https://ui.adsabs.harvard.edu/abs/2019ApJ...887...20A} {887, 20}

\bibitem[\protect\citeauthoryear{{Ashby} et~al.,}{{Ashby}
  et~al.}{2011}]{2011PASP..123.1011A}
{Ashby} M.~L.~N.,  et~al., 2011, \mn@doi [\pasp] {10.1086/661920}, \href
  {http://adsabs.harvard.edu/abs/2011PASP..123.1011A} {123, 1011}

\bibitem[\protect\citeauthoryear{{Bhattacharya} \& {van den
  Heuvel}}{{Bhattacharya} \& {van den Heuvel}}{1991}]{1991PhR...203....1B}
{Bhattacharya} D.,  {van den Heuvel} E.~P.~J.,  1991, \mn@doi [\physrep]
  {10.1016/0370-1573(91)90064-S}, \href
  {https://ui.adsabs.harvard.edu/abs/1991PhR...203....1B} {203, 1}

\bibitem[\protect\citeauthoryear{{Boquien}, {Buat}  \& {Perret}}{{Boquien}
  et~al.}{2014}]{2014A&A...571A..72B}
{Boquien} M.,  {Buat} V.,   {Perret} V.,  2014, \mn@doi [\aap]
  {10.1051/0004-6361/201424441}, \href
  {http://adsabs.harvard.edu/abs/2014A\%26A...571A..72B} {571, A72}

\bibitem[\protect\citeauthoryear{{Boquien}, {Burgarella}, {Roehlly}, {Buat},
  {Ciesla}, {Corre}, {Inoue}  \& {Salas}}{{Boquien}
  et~al.}{2019}]{2019A&A...622A.103B}
{Boquien} M.,  {Burgarella} D.,  {Roehlly} Y.,  {Buat} V.,  {Ciesla} L.,
  {Corre} D.,  {Inoue} A.~K.,   {Salas} H.,  2019, \mn@doi [\aap]
  {10.1051/0004-6361/201834156}, \href
  {https://ui.adsabs.harvard.edu/abs/2019A&A...622A.103B} {622, A103}

\bibitem[\protect\citeauthoryear{{Boroson}, {Kim}  \& {Fabbiano}}{{Boroson}
  et~al.}{2011}]{2011ApJ...729...12B}
{Boroson} B.,  {Kim} D.-W.,   {Fabbiano} G.,  2011, \mn@doi [\apj]
  {10.1088/0004-637X/729/1/12}, \href
  {https://ui.adsabs.harvard.edu/abs/2011ApJ...729...12B} {729, 12}

\bibitem[\protect\citeauthoryear{{Brorby}, {Kaaret}, {Prestwich}  \&
  {Mirabel}}{{Brorby} et~al.}{2016}]{2016MNRAS.457.4081B}
{Brorby} M.,  {Kaaret} P.,  {Prestwich} A.,   {Mirabel} I.~F.,  2016, \mn@doi
  [\mnras] {10.1093/mnras/stw284}, \href
  {https://ui.adsabs.harvard.edu/abs/2016MNRAS.457.4081B} {457, 4081}

\bibitem[\protect\citeauthoryear{{Bruzual} \& {Charlot}}{{Bruzual} \&
  {Charlot}}{2003}]{2003MNRAS.344.1000B}
{Bruzual} G.,  {Charlot} S.,  2003, \mn@doi [\mnras]
  {10.1046/j.1365-8711.2003.06897.x}, \href
  {https://ui.adsabs.harvard.edu/abs/2003MNRAS.344.1000B} {344, 1000}

\bibitem[\protect\citeauthoryear{{Cardelli}, {Clayton}  \& {Mathis}}{{Cardelli}
  et~al.}{1989}]{1989ApJ...345..245C}
{Cardelli} J.~A.,  {Clayton} G.~C.,   {Mathis} J.~S.,  1989, \mn@doi [\apj]
  {10.1086/167900}, \href
  {https://ui.adsabs.harvard.edu/abs/1989ApJ...345..245C} {345, 245}

\bibitem[\protect\citeauthoryear{{Cervi{\~n}o}, {Bongiovanni}  \&
  {Hidalgo}}{{Cervi{\~n}o} et~al.}{2016}]{2016A&A...589A.108C}
{Cervi{\~n}o} M.,  {Bongiovanni} A.,   {Hidalgo} S.,  2016, \mn@doi [\aap]
  {10.1051/0004-6361/201528056}, \href
  {http://adsabs.harvard.edu/abs/2016A\%26A...589A.108C} {589, A108}

\bibitem[\protect\citeauthoryear{{Chabrier}}{{Chabrier}}{2003}]{2003PASP..115..763C}
{Chabrier} G.,  2003, \mn@doi [\pasp] {10.1086/376392}, \href
  {https://ui.adsabs.harvard.edu/abs/2003PASP..115..763C} {115, 763}

\bibitem[\protect\citeauthoryear{{Coe}}{{Coe}}{2005}]{2005MNRAS.358.1379C}
{Coe} M.~J.,  2005, \mn@doi [\mnras] {10.1111/j.1365-2966.2005.08857.x}, \href
  {https://ui.adsabs.harvard.edu/abs/2005MNRAS.358.1379C} {358, 1379}

\bibitem[\protect\citeauthoryear{{Daddi} et~al.,}{{Daddi}
  et~al.}{2007}]{2007ApJ...670..156D}
{Daddi} E.,  et~al., 2007, \mn@doi [\apj] {10.1086/521818}, \href
  {https://ui.adsabs.harvard.edu/abs/2007ApJ...670..156D} {670, 156}

\bibitem[\protect\citeauthoryear{{Dale}, {Helou}, {Magdis}, {Armus},
  {D{\'\i}az-Santos}  \& {Shi}}{{Dale} et~al.}{2014}]{2014ApJ...784...83D}
{Dale} D.~A.,  {Helou} G.,  {Magdis} G.~E.,  {Armus} L.,  {D{\'\i}az-Santos}
  T.,   {Shi} Y.,  2014, \mn@doi [\apj] {10.1088/0004-637X/784/1/83}, \href
  {https://ui.adsabs.harvard.edu/abs/2014ApJ...784...83D} {784, 83}

\bibitem[\protect\citeauthoryear{{Davis}}{{Davis}}{2001}]{2001ApJ...548.1010D}
{Davis} J.~E.,  2001, \mn@doi [\apj] {10.1086/319002}, \href
  {https://ui.adsabs.harvard.edu/abs/2001ApJ...548.1010D} {548, 1010}

\bibitem[\protect\citeauthoryear{{Douna}, {Pellizza}, {Mirabel}  \&
  {Pedrosa}}{{Douna} et~al.}{2015}]{2015A&A...579A..44D}
{Douna} V.~M.,  {Pellizza} L.~J.,  {Mirabel} I.~F.,   {Pedrosa} S.~E.,  2015,
  \mn@doi [\aap] {10.1051/0004-6361/201525617}, \href
  {https://ui.adsabs.harvard.edu/abs/2015A&A...579A..44D} {579, A44}

\bibitem[\protect\citeauthoryear{{Draine} \& {Li}}{{Draine} \&
  {Li}}{2007}]{2007ApJ...657..810D}
{Draine} B.~T.,  {Li} A.,  2007, \mn@doi [\apj] {10.1086/511055}, \href
  {https://ui.adsabs.harvard.edu/abs/2007ApJ...657..810D} {657, 810}

\bibitem[\protect\citeauthoryear{{Draine} et~al.,}{{Draine}
  et~al.}{2014}]{2014ApJ...780..172D}
{Draine} B.~T.,  et~al., 2014, \mn@doi [\apj] {10.1088/0004-637X/780/2/172},
  \href {https://ui.adsabs.harvard.edu/abs/2014ApJ...780..172D} {780, 172}

\bibitem[\protect\citeauthoryear{{Dray}}{{Dray}}{2006}]{2006MNRAS.370.2079D}
{Dray} L.~M.,  2006, \mn@doi [\mnras] {10.1111/j.1365-2966.2006.10635.x}, \href
  {https://ui.adsabs.harvard.edu/abs/2006MNRAS.370.2079D} {370, 2079}

\bibitem[\protect\citeauthoryear{{Elbaz} et~al.,}{{Elbaz}
  et~al.}{2007}]{2007A&A...468...33E}
{Elbaz} D.,  et~al., 2007, \mn@doi [\aap] {10.1051/0004-6361:20077525}, \href
  {https://ui.adsabs.harvard.edu/abs/2007A&A...468...33E} {468, 33}

\bibitem[\protect\citeauthoryear{{Enia} et~al.,}{{Enia}
  et~al.}{2020}]{2020arXiv200104479E}
{Enia} A.,  et~al., 2020, arXiv e-prints, \href
  {https://ui.adsabs.harvard.edu/abs/2020arXiv200104479E} {p. arXiv:2001.04479}

\bibitem[\protect\citeauthoryear{{Eufrasio} et~al.,}{{Eufrasio}
  et~al.}{2017}]{2017ApJ...851...10E}
{Eufrasio} R.~T.,  et~al., 2017, \mn@doi [\apj] {10.3847/1538-4357/aa9569},
  \href {https://ui.adsabs.harvard.edu/abs/2017ApJ...851...10E} {851, 10}

\bibitem[\protect\citeauthoryear{{Fabbiano}, {Zezas}  \& {Murray}}{{Fabbiano}
  et~al.}{2001}]{2001ApJ...554.1035F}
{Fabbiano} G.,  {Zezas} A.,   {Murray} S.~S.,  2001, \mn@doi [\apj]
  {10.1086/321397}, \href
  {https://ui.adsabs.harvard.edu/abs/2001ApJ...554.1035F} {554, 1035}

\bibitem[\protect\citeauthoryear{{Foreman-Mackey}, {Hogg}, {Lang}  \&
  {Goodman}}{{Foreman-Mackey} et~al.}{2013}]{2013PASP..125..306F}
{Foreman-Mackey} D.,  {Hogg} D.~W.,  {Lang} D.,   {Goodman} J.,  2013, \mn@doi
  [\pasp] {10.1086/670067}, \href
  {https://ui.adsabs.harvard.edu/abs/2013PASP..125..306F} {125, 306}

\bibitem[\protect\citeauthoryear{{Fragos} et~al.,}{{Fragos}
  et~al.}{2013}]{2013ApJ...764...41F}
{Fragos} T.,  et~al., 2013, \mn@doi [\apj] {10.1088/0004-637X/764/1/41}, \href
  {https://ui.adsabs.harvard.edu/abs/2013ApJ...764...41F} {764, 41}

\bibitem[\protect\citeauthoryear{{Garofali}, {Williams}, {Hillis}, {Gilbert},
  {Dolphin}, {Eracleous}  \& {Binder}}{{Garofali}
  et~al.}{2018}]{2018MNRAS.479.3526G}
{Garofali} K.,  {Williams} B.~F.,  {Hillis} T.,  {Gilbert} K.~M.,  {Dolphin}
  A.~E.,  {Eracleous} M.,   {Binder} B.,  2018, \mn@doi [\mnras]
  {10.1093/mnras/sty1612}, \href
  {https://ui.adsabs.harvard.edu/abs/2018MNRAS.479.3526G} {479, 3526}

\bibitem[\protect\citeauthoryear{{Gilfanov}}{{Gilfanov}}{2004}]{2004MNRAS.349..146G}
{Gilfanov} M.,  2004, \mn@doi [\mnras] {10.1111/j.1365-2966.2004.07473.x},
  \href {https://ui.adsabs.harvard.edu/abs/2004MNRAS.349..146G} {349, 146}

\bibitem[\protect\citeauthoryear{{Gorenstein}}{{Gorenstein}}{1975}]{1975ApJ...198...95G}
{Gorenstein} P.,  1975, \mn@doi [\apj] {10.1086/153579}, \href
  {https://ui.adsabs.harvard.edu/abs/1975ApJ...198...95G} {198, 95}

\bibitem[\protect\citeauthoryear{{Grimm}, {Gilfanov}  \& {Sunyaev}}{{Grimm}
  et~al.}{2003}]{2003MNRAS.339..793G}
{Grimm} H.-J.,  {Gilfanov} M.,   {Sunyaev} R.,  2003, \mn@doi [\mnras]
  {10.1046/j.1365-8711.2003.06224.x}, \href
  {https://ui.adsabs.harvard.edu/abs/2003MNRAS.339..793G} {339, 793}

\bibitem[\protect\citeauthoryear{{Harris} \& {Zaritsky}}{{Harris} \&
  {Zaritsky}}{2009}]{2009AJ....138.1243H}
{Harris} J.,  {Zaritsky} D.,  2009, \mn@doi [\aj]
  {10.1088/0004-6256/138/5/1243}, \href
  {https://ui.adsabs.harvard.edu/abs/2009AJ....138.1243H} {138, 1243}

\bibitem[\protect\citeauthoryear{{Helou} et~al.,}{{Helou}
  et~al.}{2004}]{2004ApJS..154..253H}
{Helou} G.,  et~al., 2004, \mn@doi [\apjs] {10.1086/422640}, \href
  {http://adsabs.harvard.edu/abs/2004ApJS..154..253H} {154, 253}

\bibitem[\protect\citeauthoryear{{Jarrett}, {Cluver}, {Brown}, {Dale}, {Tsai}
  \& {Masci}}{{Jarrett} et~al.}{2019}]{2019ApJS..245...25J}
{Jarrett} T.~H.,  {Cluver} M.~E.,  {Brown} M.~J.~I.,  {Dale} D.~A.,  {Tsai}
  C.~W.,   {Masci} F.,  2019, \mn@doi [\apjs] {10.3847/1538-4365/ab521a}, \href
  {https://ui.adsabs.harvard.edu/abs/2019ApJS..245...25J} {245, 25}

\bibitem[\protect\citeauthoryear{{Kennicutt}}{{Kennicutt}}{1998}]{1998ARA&A..36..189K}
{Kennicutt} Robert~C. J.,  1998, \mn@doi [\araa]
  {10.1146/annurev.astro.36.1.189}, \href
  {https://ui.adsabs.harvard.edu/abs/1998ARA&A..36..189K} {36, 189}

\bibitem[\protect\citeauthoryear{{Kennicutt} \& {Evans}}{{Kennicutt} \&
  {Evans}}{2012}]{2012ARA&A..50..531K}
{Kennicutt} R.~C.,  {Evans} N.~J.,  2012, \mn@doi [\araa]
  {10.1146/annurev-astro-081811-125610}, \href
  {http://adsabs.harvard.edu/abs/2012ARA%26A..50..531K} {50, 531}

\bibitem[\protect\citeauthoryear{{Kennicutt}, {Lee}, {Funes}, {J.}, {Sakai}  \&
  {Akiyama}}{{Kennicutt} et~al.}{2008}]{2008ApJS..178..247K}
{Kennicutt} Jr. R.~C.,  {Lee} J.~C.,  {Funes} J.~G.,  {J.} S.,  {Sakai} S.,
  {Akiyama} S.,  2008, \mn@doi [\apjs] {10.1086/590058}, \href
  {http://adsabs.harvard.edu/abs/2008ApJS..178..247K} {178, 247}

\bibitem[\protect\citeauthoryear{{Larson} et~al.,}{{Larson}
  et~al.}{2020}]{2020ApJ...888...92L}
{Larson} K.~L.,  et~al., 2020, \mn@doi [\apj] {10.3847/1538-4357/ab5dc3}, \href
  {https://ui.adsabs.harvard.edu/abs/2020ApJ...888...92L} {888, 92}


\bibitem[\protect\citeauthoryear{{Lehmer} et~al.,}{{Lehmer}
  et~al.}{2010}]{2010ApJ...724..559L}
{Lehmer} B.~D.,  {Alexander} D.~M.,  {Bauer} F.~E.,  {Brandt} W.~N.,
  {Goulding} A.~D.,  {Jenkins} L.~P.,  {Ptak} A.,   {Roberts} T.~P.,  2010,
  \mn@doi [\apj] {10.1088/0004-637X/724/1/559}, \href
  {https://ui.adsabs.harvard.edu/abs/2010ApJ...724..559L} {724, 559}

\bibitem[\protect\citeauthoryear{{Lehmer} et~al.,}{{Lehmer}
  et~al.}{2016}]{2016ApJ...825....7L}
{Lehmer} B.~D.,  et~al., 2016, \mn@doi [\apj] {10.3847/0004-637X/825/1/7},
  \href {https://ui.adsabs.harvard.edu/abs/2016ApJ...825....7L} {825, 7}

\bibitem[\protect\citeauthoryear{{Lehmer} et~al.,}{{Lehmer}
  et~al.}{2019}]{2019ApJS..243....3L}
{Lehmer} B.~D.,  et~al., 2019, \mn@doi [\apjs] {10.3847/1538-4365/ab22a8},
  \href {https://ui.adsabs.harvard.edu/abs/2019ApJS..243....3L} {243, 3}

\bibitem[\protect\citeauthoryear{{Linden}, {Kalogera}, {Sepinsky}, {Prestwich},
  {Zezas}  \& {Gallagher}}{{Linden} et~al.}{2010}]{2010ApJ...725.1984L}
{Linden} T.,  {Kalogera} V.,  {Sepinsky} J.~F.,  {Prestwich} A.,  {Zezas} A.,
  {Gallagher} J.~S.,  2010, \mn@doi [\apj] {10.1088/0004-637X/725/2/1984},
  \href {https://ui.adsabs.harvard.edu/abs/2010ApJ...725.1984L} {725, 1984}

\bibitem[\protect\citeauthoryear{{Mahajan}, {Ashby}, {Willner}, {Barmby},
  {Fazio}, {Maragkoudakis}, {Raychaudhury}  \& {Zezas}}{{Mahajan}
  et~al.}{2019}]{2019MNRAS.482..560M}
{Mahajan} S.,  {Ashby} M.~L.~N.,  {Willner} S.~P.,  {Barmby} P.,  {Fazio}
  G.~G.,  {Maragkoudakis} A.,  {Raychaudhury} S.,   {Zezas} A.,  2019, \mn@doi
  [\mnras] {10.1093/mnras/sty2699}, \href
  {https://ui.adsabs.harvard.edu/abs/2019MNRAS.482..560M} {482, 560}

\bibitem[\protect\citeauthoryear{{Mapelli}, {Colpi}  \& {Zampieri}}{{Mapelli}
  et~al.}{2009}]{2009MNRAS.395L..71M}
{Mapelli} M.,  {Colpi} M.,   {Zampieri} L.,  2009, \mn@doi [\mnras]
  {10.1111/j.1745-3933.2009.00645.x}, \href
  {https://ui.adsabs.harvard.edu/abs/2009MNRAS.395L..71M} {395, L71}

\bibitem[\protect\citeauthoryear{{Mapelli}, {Ripamonti}, {Zampieri}, {Colpi}
  \& {Bressan}}{{Mapelli} et~al.}{2010}]{2010MNRAS.408..234M}
{Mapelli} M.,  {Ripamonti} E.,  {Zampieri} L.,  {Colpi} M.,   {Bressan} A.,
  2010, \mn@doi [\mnras] {10.1111/j.1365-2966.2010.17048.x}, \href
  {https://ui.adsabs.harvard.edu/abs/2010MNRAS.408..234M} {408, 234}

\bibitem[\protect\citeauthoryear{{Maragkoudakis}, {Zezas}, {Ashby}  \&
  {Willner}}{{Maragkoudakis} et~al.}{2017}]{2017MNRAS.466.1192M}
{Maragkoudakis} A.,  {Zezas} A.,  {Ashby} M.~L.~N.,   {Willner} S.~P.,  2017,
  \mn@doi [\mnras] {10.1093/mnras/stw3180}, \href
  {http://adsabs.harvard.edu/abs/2017MNRAS.466.1192M} {466, 1192}

\bibitem[\protect\citeauthoryear{{Maragkoudakis}, {Zezas}, {Ashby}  \&
  {Willner}}{{Maragkoudakis} et~al.}{2018}]{2018MNRAS.475.1485M}
{Maragkoudakis} A.,  {Zezas} A.,  {Ashby} M.~L.~N.,   {Willner} S.~P.,  2018,
  \mn@doi [\mnras] {10.1093/mnras/stx3247}, \href
  {https://ui.adsabs.harvard.edu/abs/2018MNRAS.475.1485M} {475, 1485}

\bibitem[\protect\citeauthoryear{{Massey}, {Strobel}, {Barnes}  \&
  {Anderson}}{{Massey} et~al.}{1988}]{1988ApJ...328..315M}
{Massey} P.,  {Strobel} K.,  {Barnes} J.~V.,   {Anderson} E.,  1988, \mn@doi
  [\apj] {10.1086/166294}, \href
  {https://ui.adsabs.harvard.edu/abs/1988ApJ...328..315M} {328, 315}

\bibitem[\protect\citeauthoryear{{Mineo}, {Gilfanov}  \& {Sunyaev}}{{Mineo}
  et~al.}{2012a}]{2012MNRAS.419.2095M}
{Mineo} S.,  {Gilfanov} M.,   {Sunyaev} R.,  2012a, \mn@doi [\mnras]
  {10.1111/j.1365-2966.2011.19862.x}, \href
  {https://ui.adsabs.harvard.edu/abs/2012MNRAS.419.2095M} {419, 2095}

\bibitem[\protect\citeauthoryear{{Mineo}, {Gilfanov}  \& {Sunyaev}}{{Mineo}
  et~al.}{2012b}]{2012MNRAS.426.1870M}
{Mineo} S.,  {Gilfanov} M.,   {Sunyaev} R.,  2012b, \mn@doi [\mnras]
  {10.1111/j.1365-2966.2012.21831.x}, \href
  {https://ui.adsabs.harvard.edu/abs/2012MNRAS.426.1870M} {426, 1870}

\bibitem[\protect\citeauthoryear{{Mineo}, {Gilfanov}, {Lehmer}, {Morrison}  \&
  {Sunyaev}}{{Mineo} et~al.}{2014}]{2014MNRAS.437.1698M}
{Mineo} S.,  {Gilfanov} M.,  {Lehmer} B.~D.,  {Morrison} G.~E.,   {Sunyaev} R.,
   2014, \mn@doi [\mnras] {10.1093/mnras/stt1999}, \href
  {https://ui.adsabs.harvard.edu/abs/2014MNRAS.437.1698M} {437, 1698}

\bibitem[\protect\citeauthoryear{{Morrison} \& {McCammon}}{{Morrison} \&
  {McCammon}}{1983}]{1983ApJ...270..119M}
{Morrison} R.,  {McCammon} D.,  1983, \mn@doi [\apj] {10.1086/161102}, \href
  {https://ui.adsabs.harvard.edu/abs/1983ApJ...270..119M} {270, 119}

\bibitem[\protect\citeauthoryear{{Murphy} et~al.,}{{Murphy}
  et~al.}{2011}]{2011ApJ...737...67M}
{Murphy} E.~J.,  et~al., 2011, \mn@doi [\apj] {10.1088/0004-637X/737/2/67},
  \href {http://adsabs.harvard.edu/abs/2011ApJ...737...67M} {737, 67}

\bibitem[\protect\citeauthoryear{{Noeske} et~al.,}{{Noeske}
  et~al.}{2007}]{2007ApJ...660L..47N}
{Noeske} K.~G.,  et~al., 2007, \mn@doi [\apjl] {10.1086/517927}, \href
  {https://ui.adsabs.harvard.edu/abs/2007ApJ...660L..47N} {660, L47}

\bibitem[\protect\citeauthoryear{{Osterbrock} \& {Ferland}}{{Osterbrock} \&
  {Ferland}}{2006}]{2006agna.book.....O}
{Osterbrock} D.~E.,  {Ferland} G.~J.,  2006, {Astrophysics of gaseous nebulae
  and active galactic nuclei}

\bibitem[\protect\citeauthoryear{{Pancoast}, {Sajina}, {Lacy}, {Noriega-Crespo}
   \& {Rho}}{{Pancoast} et~al.}{2010}]{2010ApJ...723..530P}
{Pancoast} A.,  {Sajina} A.,  {Lacy} M.,  {Noriega-Crespo} A.,   {Rho} J.,
  2010, \mn@doi [\apj] {10.1088/0004-637X/723/1/530}, \href
  {http://adsabs.harvard.edu/abs/2010ApJ...723..530P} {723, 530}

\bibitem[\protect\citeauthoryear{{Park}, {Kashyap}, {Siemiginowska}, {van Dyk},
  {Zezas}, {Heinke}  \& {Wargelin}}{{Park} et~al.}{2006}]{2006ApJ...652..610P}
{Park} T.,  {Kashyap} V.~L.,  {Siemiginowska} A.,  {van Dyk} D.~A.,  {Zezas}
  A.,  {Heinke} C.,   {Wargelin} B.~J.,  2006, \mn@doi [\apj] {10.1086/507406},
  \href {https://ui.adsabs.harvard.edu/abs/2006ApJ...652..610P} {652, 610}

\bibitem[\protect\citeauthoryear{{Peeters}, {Spoon}  \& {Tielens}}{{Peeters}
  et~al.}{2004}]{2004ApJ...613..986P}
{Peeters} E.,  {Spoon} H.~W.~W.,   {Tielens} A.~G.~G.~M.,  2004, \mn@doi [\apj]
  {10.1086/423237}, \href
  {https://ui.adsabs.harvard.edu/abs/2004ApJ...613..986P} {613, 986}

\bibitem[\protect\citeauthoryear{{Podsiadlowski}, {Langer}, {Poelarends},
  {Rappaport}, {Heger}  \& {Pfahl}}{{Podsiadlowski} et~al.}{2004}]{Podsi2004}
{Podsiadlowski} P.,  {Langer} N.,  {Poelarends} A.~J.~T.,  {Rappaport} S.,
  {Heger} A.,   {Pfahl} E.,  2004, \mn@doi [\apj] {10.1086/421713}, \href
  {https://ui.adsabs.harvard.edu/abs/2004ApJ...612.1044P} {612, 1044}

\bibitem[\protect\citeauthoryear{{Politakis}, {Zezas}, {Andrews}  \&
  {Williams}}{{Politakis} et~al.}{2020}]{2020MNRAS.tmp..540P}
{Politakis} B.,  {Zezas} A.,  {Andrews} J.~J.,   {Williams} S.~J.,  2020,
  \mn@doi [\mnras] {10.1093/mnras/staa561}, \href
  {https://ui.adsabs.harvard.edu/abs/2020MNRAS.tmp..540P} {}

\bibitem[\protect\citeauthoryear{{Prestwich}, {Tsantaki}, {Zezas}, {Jackson},
  {Roberts}, {Foltz}, {Linden}  \& {Kalogera}}{{Prestwich}
  et~al.}{2013}]{2013ApJ...769...92P}
{Prestwich} A.~H.,  {Tsantaki} M.,  {Zezas} A.,  {Jackson} F.,  {Roberts}
  T.~P.,  {Foltz} R.,  {Linden} T.,   {Kalogera} V.,  2013, \mn@doi [\apj]
  {10.1088/0004-637X/769/2/92}, \href
  {https://ui.adsabs.harvard.edu/abs/2013ApJ...769...92P} {769, 92}

\bibitem[\protect\citeauthoryear{{Ranalli}, {Comastri}  \& {Setti}}{{Ranalli}
  et~al.}{2003}]{2003A&A...399...39R}
{Ranalli} P.,  {Comastri} A.,   {Setti} G.,  2003, \mn@doi [\aap]
  {10.1051/0004-6361:20021600}, \href
  {https://ui.adsabs.harvard.edu/abs/2003A%26A...399...39R} {399, 39}

\bibitem[\protect\citeauthoryear{{Rieke}, {Alonso-Herrero}, {Weiner},
  {P{\'e}rez-Gonz{\'a}lez}, {Blaylock}, {Donley}  \& {Marcillac}}{{Rieke}
  et~al.}{2009}]{2009ApJ...692..556R}
{Rieke} G.~H.,  {Alonso-Herrero} A.,  {Weiner} B.~J.,  {P{\'e}rez-Gonz{\'a}lez}
  P.~G.,  {Blaylock} M.,  {Donley} J.~L.,   {Marcillac} D.,  2009, \mn@doi
  [\apj] {10.1088/0004-637X/692/1/556}, \href
  {http://adsabs.harvard.edu/abs/2009ApJ...692..556R} {692, 556}

\bibitem[\protect\citeauthoryear{{Salpeter}}{{Salpeter}}{1955}]{1955ApJ...121..161S}
{Salpeter} E.~E.,  1955, \mn@doi [\apj] {10.1086/145971}, \href
  {https://ui.adsabs.harvard.edu/abs/1955ApJ...121..161S} {121, 161}

\bibitem[\protect\citeauthoryear{{Smith}, {Brickhouse}, {Liedahl}  \&
  {Raymond}}{{Smith} et~al.}{2001}]{2001ApJ...556L..91S}
{Smith} R.~K.,  {Brickhouse} N.~S.,  {Liedahl} D.~A.,   {Raymond} J.~C.,  2001,
  \mn@doi [\apjl] {10.1086/322992}, \href
  {https://ui.adsabs.harvard.edu/abs/2001ApJ...556L..91S} {556, L91}

\bibitem[\protect\citeauthoryear{{Tauris} \& {van den Heuvel}}{{Tauris} \& {van
  den Heuvel}}{2006}]{2006csxs.book..623T}
{Tauris} T.~M.,  {van den Heuvel} E.~P.~J.,  2006, {Formation and evolution of
  compact stellar X-ray sources}.
pp 623--665

\bibitem[\protect\citeauthoryear{{Xiang} et~al.,}{{Xiang}
  et~al.}{2018}]{2018ApJS..237...33X}
{Xiang} M.,  et~al., 2018, \mn@doi [\apjs] {10.3847/1538-4365/aad237}, \href
  {https://ui.adsabs.harvard.edu/abs/2018ApJS..237...33X} {237, 33}

\bibitem[\protect\citeauthoryear{{Zezas}, {Fabbiano}, {Baldi}, {Schweizer},
  {King}, {Ponman}  \& {Rots}}{{Zezas} et~al.}{2006}]{2006ApJS..166..211Z}
{Zezas} A.,  {Fabbiano} G.,  {Baldi} A.,  {Schweizer} F.,  {King} A.~R.,
  {Ponman} T.~J.,   {Rots} A.~H.,  2006, \mn@doi [\apjs] {10.1086/501526},
  \href
  {https://ui-adsabs-harvard-edu.ezp-prod1.hul.harvard.edu/abs/2006ApJS..166..211Z}
  {166, 211}

\bibitem[\protect\citeauthoryear{{Zhang}, {Gilfanov}  \& {Bogd{\'a}n}}{{Zhang}
  et~al.}{2012}]{2012A&A...546A..36Z}
{Zhang} Z.,  {Gilfanov} M.,   {Bogd{\'a}n} {\'A}.,  2012, \mn@doi [\aap]
  {10.1051/0004-6361/201219015}, \href
  {https://ui.adsabs.harvard.edu/abs/2012A&A...546A..36Z} {546, A36}

\bibitem[\protect\citeauthoryear{{Zhu}, {Wu}, {Li}  \& {Cao}}{{Zhu}
  et~al.}{2010}]{2010RAA....10..329Z}
{Zhu} Y.-N.,  {Wu} H.,  {Li} H.-N.,   {Cao} C.,  2010, \mn@doi [Research in
  Astronomy and Astrophysics] {10.1088/1674-4527/10/4/004}, \href
  {http://adsabs.harvard.edu/abs/2010RAA....10..329Z} {10, 329}

\bibitem[\protect\citeauthoryear{{van Dyk}, {Connors}, {Kashyap}  \&
  {Siemiginowska}}{{van Dyk} et~al.}{2001}]{2001ApJ...548..224V}
{van Dyk} D.~A.,  {Connors} A.,  {Kashyap} V.~L.,   {Siemiginowska} A.,  2001,
  \mn@doi [\apj] {10.1086/318656}, \href
  {https://ui.adsabs.harvard.edu/abs/2001ApJ...548..224V} {548, 224}

\bibitem[\protect\citeauthoryear{{van den Heuvel}, {Portegies Zwart},
  {Bhattacharya}  \& {Kaper}}{{van den Heuvel}
  et~al.}{2000}]{2000A&A...364..563V}
{van den Heuvel} E.~P.~J.,  {Portegies Zwart} S.~F.,  {Bhattacharya} D.,
  {Kaper} L.,  2000, \aap, \href
  {https://ui.adsabs.harvard.edu/abs/2000A&A...364..563V} {364, 563}

\end{thebibliography}




\appendix

\section{Maximum likelihood fitting method}
\label{sec:Apendix_Likelihood}

We fit a linear model with intrinsic scatter to the SFR and X-ray
luminosity of the regions of all galaxies. Specifically, we consider
the errors-in-variables regression model:
\begin{align}
    \begin{split}
        x_i &= x_i^t + \eta_i \\
        y_i &= y_i^t + \zeta_i \\
        y_i^t &= a x_i^t + b + \epsilon\left(x_i^t\right) \quad,
    \end{split}
\end{align}
where $x_i$ and $y_i$ are the observed $\log{\rm SFR}$ and
$\log{L_X}$ of the $i$-th region, while $x_i^t$ and $y_i^t$ are the
respective intrinsic values; $\eta_i$ and $\zeta_i$ denote the error
distributions on $x_i$ and $y_i$ respectively.\footnote{$\eta_i$ and
  $\zeta_i$ are not normally distributed because they represent the
  logarithmic transformation of the truncated Gaussian errors on SFRs
  (zero-truncated) and the logarithm of the X-ray
  luminosity (Section \ref{sec:Sub-galactic analysis})} For the
intrinsic scatter $\epsilon$ we consider two cases: (i) constant:
\begin{equation}
    \epsilon\left(x_i^t\right) = \sigma\quad, \mbox{ where $\sigma {\ge} 0$}
\end{equation}
and (ii) including a term linear in the logarithm of SFR:
\begin{equation}
    \epsilon\left(x_i^t\right) = \max\left\{0, \sigma_1 x_i^t + \sigma_2\right\}
\end{equation}
where the \lq{}$\max$\rq{} function ensures that the intrinsic
scatter is non-negative. 

Assuming independent measurements, the posterior probability of the
model parameters, 
$\vec{p} = \left(a, b, \sigma\right)$ or
$\left(a, b, \sigma_1, \sigma_2\right)$:
\begin{equation}
    \pi(\vec{p}) \prod\limits_{i} P(x_i, y_i | \vec{p}) \quad,
\end{equation}
where the prior is the product of the priors of each parameter
\begin{equation}
    \pi(\vec{p}) = 
    \pi(a) \pi(b) \pi(\sigma) \quad \mbox{or} \quad
    \pi(a) \pi(b) \pi(\sigma_1) \pi(\sigma_2) \quad,
\end{equation}
and the datum likelihood is the marginalization of the likelihood
considering all possible values for the intrinsic SFR and X-ray
luminosity
\begin{equation}
    P(x_i, y_i | \vec{p}) = 
        \iint
        P(x_i, y_i, x_i^t, y_i^t | \vec{p})\,dx_i^t\,dy_i^t \quad .
        \label{eq:datumlikelihood}
\end{equation}
Considering that (i) the observed values depend only on the
measurement errors and the intrinsic values, (ii) the intrinsic
values depend only on the intrinsic model, and (iii) the errors on
$x_i$ and $y_i$ are independent, the integrand of
\eqref{eq:datumlikelihood} becomes
\begin{equation}
    P\left(x_i | x_i^t, \eta_i^t\right)
    P\left(y_i | y_i^t, \zeta_i^t\right)
    P(y_i^t|x_i^t, \vec{p}) P(x_i^t|\vec{p}) \quad,
    \label{eq:integrand}
\end{equation}
where the probability of $x_i$ and $y_i$ was computed using the
corresponding distributions of $\eta_i$ and $\zeta_i$, the prior on
$x_i^t$ was chosen to be uniform between two bounds $x_{\min{}}^t$
and $x_{\max{}}^t$ (ensuring that they enclose all the observed
values $x_i$ and $3\sigma$ around them), and the probability of
$y_i^t$ was given by the normal distribution density considering the
intrinsic mean and scatter:
\begin{equation}
    \Bigg(\frac{1}{{2\pi \epsilon^2\left(x_i^t\right)}}\Bigg)^{1/2}
    {\rm exp}\left[
        -\frac{\left(y_i^t - a x_i^t - b\right)^2}
        {2\epsilon^2\left(x_i^t\right)} 
    \right] \quad.
\end{equation}

The model parameters $a$, $b$ and $\sigma$ (or $\sigma_1$ and
$\sigma_2$) were estimated by sampling the posterior distribution
using the Markov Chain Monte Carlo technique. Specifically, we used
the \texttt{emcee 3.0rc2} package for Python 3
\citep{2013PASP..125..306F} with uniform priors for the model
parameters, sufficiently wide to not be very informative but
narrow enough to aid the convergence of the MCMC chains, i.e.,
$
        a \in \left[0, 2\right], 
        b \in \left[38, 41\right], 
        \sigma \in \left[0, 2\right],
        \sigma_1 \in \left[-1, 1\right]
        \mbox{ and } 
        \sigma_2 \in \left[0, 2\right]
$.
The priors were also used to sample the initial positions of the
Markov chains.

In order to fit the scaling with both the SFR and the stellar mass, i.e.,
\begin{equation}
    \rm{log} \, \textit{L}_X = \rm{log}(10^{\alpha + log \, \rm{SFR}}
    + 10^{\beta + log \, M_\star}) + \sigma \quad,
\end{equation}
we employed the intrinsic mean model
\begin{equation}
    y_i^t = \log\left(10^{\alpha + x_i^t} + 10^{\beta + m_i^t}\right) \quad,
\end{equation}
where $m_i^t$ is the logarithm of the stellar mass of the $i$-th
region with error distribution $\xi_i$ with respect to its intrinsic
value:
\begin{equation}
    m_i = m_i^t + \xi_i \quad.
\end{equation}
Now, the datum likelihood is a triple integral,
\begin{equation}
    P(x_i, m_i, y_i | \vec{p}) = \iiint P\left(x_i, m_i, y_i, x_i^t,
      m_i^t, y_i^t|\vec{p}\right) \quad, 
\end{equation}
but using the same assumptions as before (i.e., independent
measurements), the integral is the same as in equation
\ref{eq:integrand} with an additional multiplicative PDF term for the stellar mass measurement
$P\left(m_i|m_i^t,\xi_i\right)$.

\begin{figure}
\centering
\includegraphics[width=\columnwidth]{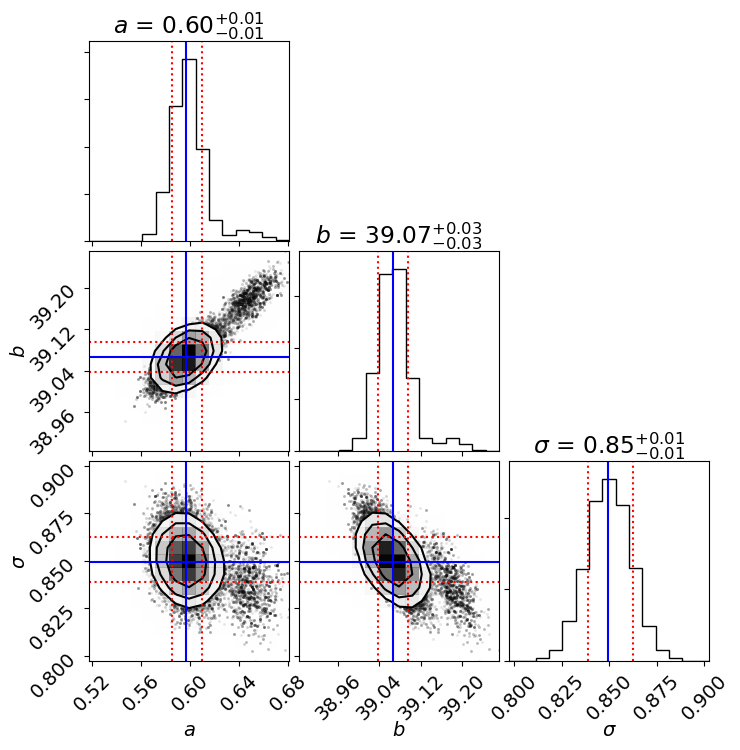}
    \caption{The marginal posterior distributions of the three
      parameters of the model: $\rm{\log} \textit{L}_X = \textit{a}
      \, \rm{\log}SFR + \textit{b} + \sigma$ in the case of $\rm
      H\alpha$ SFR and $1 {\times} 1 \, \rm{kpc}^2$ sub-galactic
      regions. }
    \label{fig:corner_H1}
\end{figure}

Results are shown in Figures~\ref{fig:Fit_Lx_SFR_all},
\ref{fig:Fit_Lx_SFR_TSH}, \ref{fig:Lx_SFR_separ}, and
\ref{fig:LxSFR_sSFR} and Tables~\ref{tab:Fit_results_2},
\ref{tab:Fit_results}, and \ref{tab:Fit_results_2D}. An example of
the results of the fits is shown in Fig.~\ref{fig:corner_H1}.

\bsp	
\label{lastpage}
\end{document}